# The nature of giant clumps in distant galaxies probed by the anatomy of the Cosmic Snake

Antonio Cava[1], Daniel Schaerer[1,2], Johan Richard[3], Pablo G. Pérez-González[4], Miroslava Dessauges-Zavadsky[1], Lucio Mayer[5,6] & Valentina Tamburello[5,6]

[1]Observatoire de Genève, Université de Genève, 51 Ch. des Maillettes, 1290 Versoix, Switzerland

[2]CNRS, IRAP, 14 Avenue E. Belin, 31400 Toulouse, France

[3]Univ Lyon, Univ Lyon1, Ens de Lyon, CNRS, Centre de Recherche Astrophysique de Lyon, UMR5574, F-69230, Saint-Genis-Laval, France

[4]Departamento de Astrofísica y Ciencias de la Atmósfera, Facultad de CC. Físicas, Universidad Complutense de Madrid, E-28040, Madrid, Spain

[5]Center for Theoretical Astrophysics and Cosmology, Institute for Computational Science, University of Zurich, Winterthurerstrasse 190, CH-8057 Zürich, Switzerland

[6]Physik-Institut, University of Zurich, Winterthurerstrasse 190, CH-8057 Zürich, Switzerland

**Giant stellar clumps are ubiquitous in high-redshift galaxies.[1,2] They are thought to play an important role in the build-up of galactic bulges[3] and as diagnostics of star formation feedback in galactic discs.[4] Hubble Space Telescope (HST) blank field imaging surveys have estimated that these clumps have masses up to 10$^{9.5}$M$_\odot$ and linear sizes $\gtrsim 1$ kpc.[5,6] Recently, gravitational lensing has also been used to get higher spatial resolution.[7–9] However, both recent lensed observations[10,11] and models[12,13] suggest that the clumps properties may be overestimated by the limited resolution of standard imaging techniques. A definitive proof of this observational bias is nevertheless still missing. Here we investigate**

**directly the effect of resolution on clump properties by analysing multiple gravitationally-lensed images of the same galaxy at different spatial resolutions, down to 30 pc. We show that the typical mass and size of giant clumps, generally observed at $\sim$1 kpc resolution in high-redshift galaxies, are systematically overestimated. The high spatial resolution data, only enabled by strong gravitational lensing using currently available facilities, support smaller scales of clump formation by fragmentation of the galactic gas disk via gravitational instabilities.**

The multiply imaged galaxy situated behind the central region of the cluster MACSJ1206.2-084747[14] (Figure 1a) is the perfect target for our experiment. Figure 1b shows two images of this galaxy: a strongly lensed one whose peculiar shape led us to name it the ”Cosmic Snake”, and a more regular and less amplified one (the ”Counterimage”). For the source galaxy we have estimated, from spectral energy distribution (SED) fitting (see Methods), a total stellar mass $M_* \sim 4\times10^{10} M_\odot$, and total SFR$\sim 30 M_\odot\,yr^{-1}$ This places the galaxy on the main sequence of star-forming galaxies at z$\sim$1–2, providing a target with physical properties comparable to typical high redshift clumpy galaxies.[15,16]

According to our tailored lensing model (see Methods) the Cosmic Snake is composed by four elongated and stretched images (Figure 1c) of the southern half of the source galaxy, imaged with magnifications covering a wide range from a few to hundreds times (Figure 1d). In contrast, the Counterimage is located in a region of nearly constant amplification, with an average magnification of $\mu \simeq 4.5$, and shows the entire source galaxy. These extremely different magnifications allow us to inspect features of the galaxy on very different intrinsic physical scales.

Using high-quality data from the Cluster Lensing And Supernova survey with Hubble (CLASH; *17*), we have identified clumps in the Cosmic Snake and the Counterimage starting from the rest-frame UV band (HST-WFC3 filter F390W). This filter shows clumps with a high contrast and attains the best spatial resolution available ($\mathrm{FWHM_{F390W}}$ $\sim$0.07$''$). We complement this sample by selecting clumps in the rest-frame optical band (HST-WFC3 filter F110W, $\mathrm{FWHM_{F110W}}$ $\sim$0.13$''$). We refer to the two selections as "blue" and "red" clumps respectively, as the latter have redder SEDs, by definition. We identify a total of 79 clumps, 24 in the Counterimage and 55 in the Cosmic Snake (19 in the Northern region, 19 in the Southern region, 10 in the central-Northern region, 7 central-Southern region; see, Figure 1c). Globally, we have 21 red clumps and 58 blue clumps. We define the size of the clumps as the circularized radius (including PSF correction) corresponding to the elliptical region of each clump. These regions are carefully defined based on multiple visual and automatic (isophotes and/or 2D-gaussian fits) criteria. The full SED is extracted from the 16 available HST bands and fitted using the *Hyperz* fitting code.[18] This fitting code provide us with the physical parameters (principally stellar mass) of the clumps. See Methods section for details about the clump selection, photometric extraction, and SED fitting.

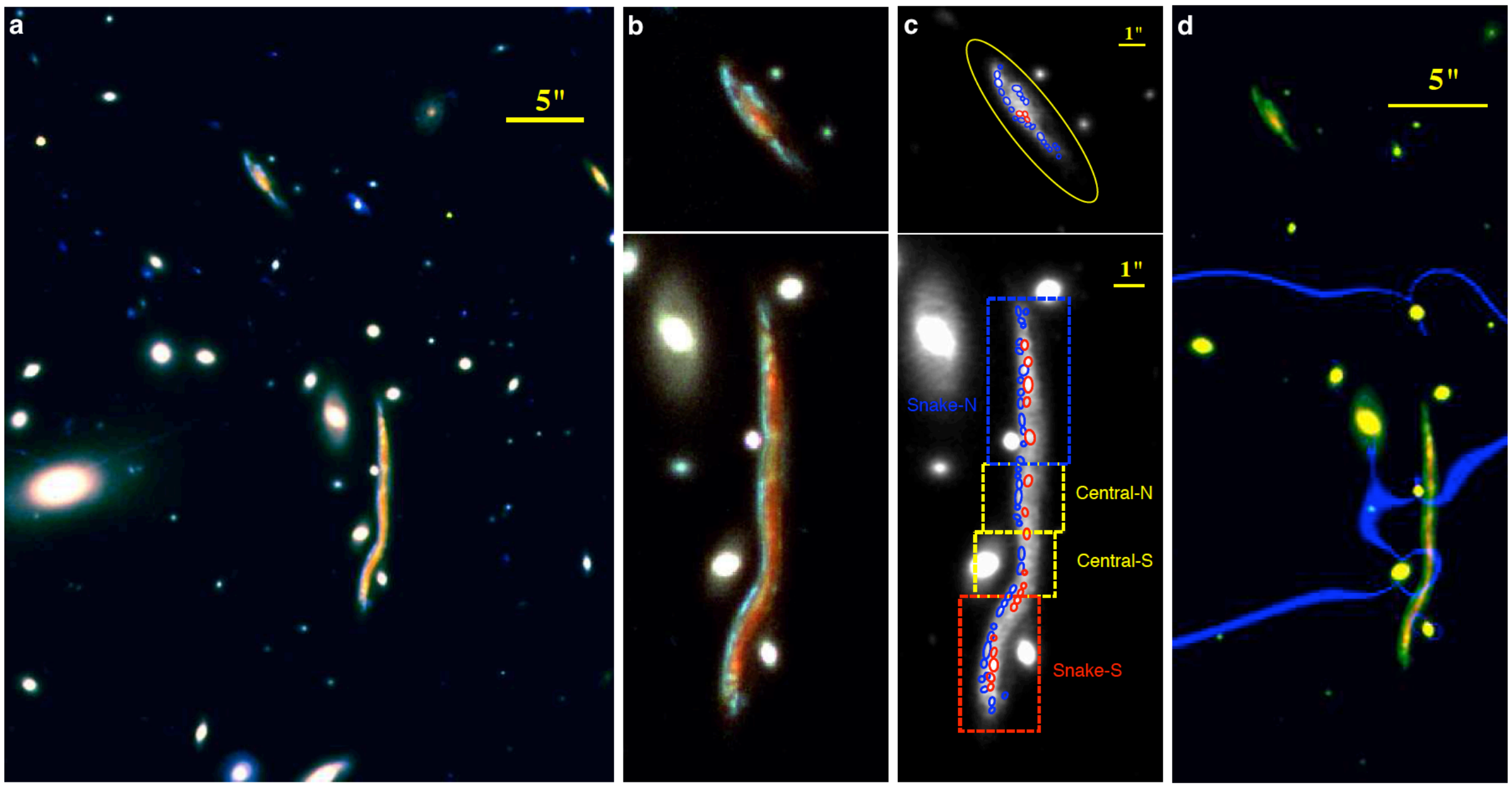

**Figure 1 | Overview of the *Cosmic Snake* and the *Counterimage*.** **a**, Portion of the HST field of view showing an RGB color composite image (R = F160W, G = F110W, B =F606W) of the cluster MACSJ1206.2-0847 including the giant arc (*Cosmic Snake*) and its *Counterimage*. **b**, Zoomed view of the Cosmic Snake (bottom) and the Counterimage(top). **c**, Image of the Cosmic Snake (bottom) and its Counterimage (top) with regions defined as clumps (blue regions for blue clumps, red regions for red clumps, yellow for whole galaxy). Rectangular areas define the four portions of the Cosmic Snake (corresponding to multiple images). **d**, RGB composite image including: R = F160W, G = F110W, B = amplification map. For the fiducial lensing model, blue shaded areas indicate amplification above 100, close to the critical lines. Representative scale bars are provided in each panel (same scale for panels **b** and **c**).

In Figure 2a we show the mass-size diagram of all the clumps. From this plot we clearly see the effect of amplification on the Cosmic Snake, which allows us to attain a physical scale of $\sim 30$ parsecs (vertical dot-dashed line) in regions with the highest magnification. The Counterimage is much less amplified, so our intrinsic resolution is limited to $\sim 300$ parsecs scale (vertical dashed line). Interestingly the clumps observed in the Counterimage are clearly larger and more massive (respectively, by a factor of 2-3 and 4-5 on average, with typical masses $\gtrsim 10^8\,M_\odot$) than clumps in the Cosmic Snake, where we find masses down to $\sim 10^7\,M_\odot$. The clumps in the Counterimage appear comparable to what has been claimed in previous studies of giant clumps in high-redshift field (i.e., not amplified) galaxies.[5,15]

In Figure 2b we show the clump mass distribution for clumps in the Cosmic Snake (orange filled histogram) and the Counterimage (green shaded histogram). Since both the Cosmic Snake and the Counterimage are multiple images of the same source galaxy, the observed size and mass shift provides direct proof of the artificial boosting of the inferred clump properties in lower resolution and less deep imaging, as claimed in our recent work,[10] where we have shown that clumps observed in strongly lensed galaxies are characterized by smaller masses, with respect to the clumps detected in unlensed galaxies. Higher resolutions and depths, enabled by strong gravitational lensing, clearly reveal that clumps have intrinsically smaller masses. This finding is also in line with recent computational studies, which examine the formation of gaseous and stellar clumps in high-resolution simulations.[12,13,19] Both the average and shape of the clump mass distribution from the Cosmic Snake are in qualitative agreement with the simulated clumps,[19] suggesting that simulations and observations start to converge towards a coherent view.

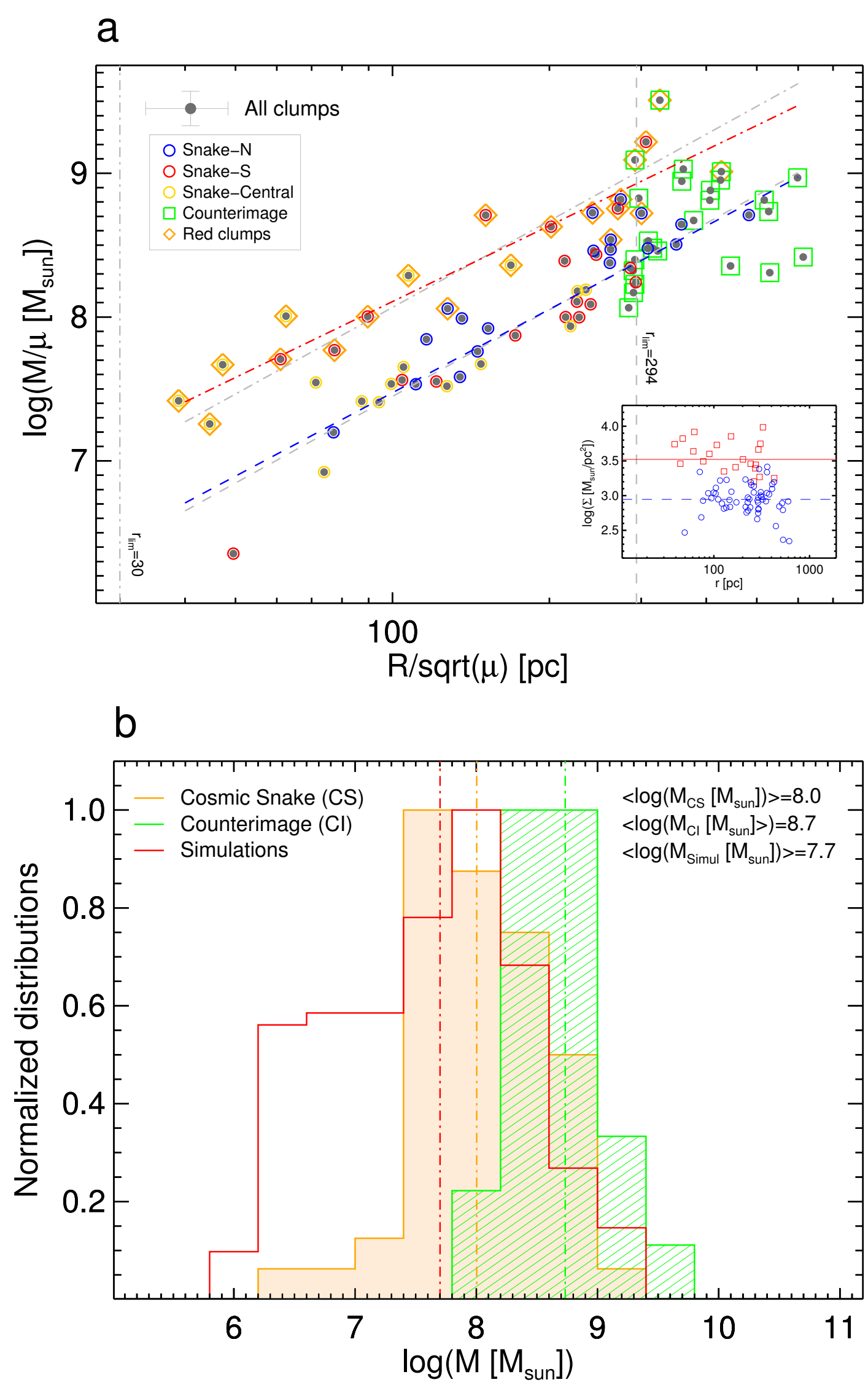

**Figure 2 | Intrinsic mass and size of the clumps, corrected for lensing.** **a**, Mass-size diagram (top panel) for the Cosmic Snake and the Counterimage ($\mu$ is the amplification factor). The vertical lines represent our resolution limit for the amplification of the Counterimage (dashed grey line) and for the best resolution reacheable in the Cosmic Snake (dot-dashed grey line). The blue and red dashed-lines are the fit to the mass-size relations for blue and red clumps, respectively. For comparison, we also overplot two constant surface density relations (grey lines). The inset shows the density-size plot for blue (circles) and red (squares) clumps. The point with error bars above the legend provides typical uncertainties. **b**, Clump mass distribution from this work (filled orange and dashed green histograms for the Cosmic Snake and Counterimage clumps respectively), compared to stellar clumps from the high-resolution simulations of Tamburello et al. (2015) for z∼2 galaxies. Vertical lines represent the median values, also reported in the legend.

Another interesting feature apparent in Figure 2a is that both the blue and red clumps follow a tight relation between stellar mass and radius (although shifted by $\sim$0.6 dex in mass), with a slope close to $M \propto r^2$, which corresponds to a constant surface density. A similar relation has been found for the giant molecular clouds of the Milky Way and nearby galaxies.[20] The corresponding derived stellar mass densities are high, between $\sim 10^{2.5}$ and $10^4\,\mathrm{M_\odot\,pc^{-2}}$ (see inset in Figure 2a), typical of globular clusters and super star clusters in local galaxies.[21] For each (red or blue) clump sample the lower limit in the mass-radius relation is due to a selection threshold in the respective filter, corresponding to a constant surface brightness. However, the fact that the maximum stellar surface density of blue clumps is limited and lower than that of red clumps, implies that the red clumps tend to be denser, on average, than blue clumps.

To understand why the red clumps are the densest, and examine possible implications, we have investigated the properties of the clumps as a function of the position within the host galaxy. We find a clear trend in the mass-galactocentric distance distribution shown in Figure 3, with red clumps occupying the central region of the galaxy and the maximum clump mass decreasing with distance. Possible systematic effects introduced by galaxy background variations are discussed in detail in the Methods section, but we stress here that the strong radial dependence of clumps mass cannot be mimicked by these effects (at most it could introduce a shallower slope). Indeed, the trend shown in Figure 3 is applicable also taking into account only the upper envelope of the mass-distance distribution for each part of the Snake, or the CI. This implies that the trend cannot be driven by a flux detection bias, which wold not affect the most massive clumps.

These observational findings altogether favour the following picture, already suggested also in previous works[3,5,22,23] : clumps form by fragmentation from a turbulent galactic gas disk due to gravitational instability and start to slowly spiral toward the centre of the galaxy. During their orbital evolution, the clumps are affected by interactions with the surrounding material which generate strong dynamical friction (merging/collision with other clumps, mass accretion from diffuse gas). The final fate of the original clumps can be accretion to the central core of the galaxy or dissolution due to dynamical processes and feedback mechanisms.[3,4,24–26] The fact that the red clumps are the densest and close to the center, would be a consequence of the fact that only the densest clumps survive the dynamical interactions within the galaxy disk. Although age determinations of the clumps are very uncertain and depend strongly on their unknown star formation histories, we find some tentative evidence for older ages of the red clumps compared to the blue ones, supporting also the above depicted picture.

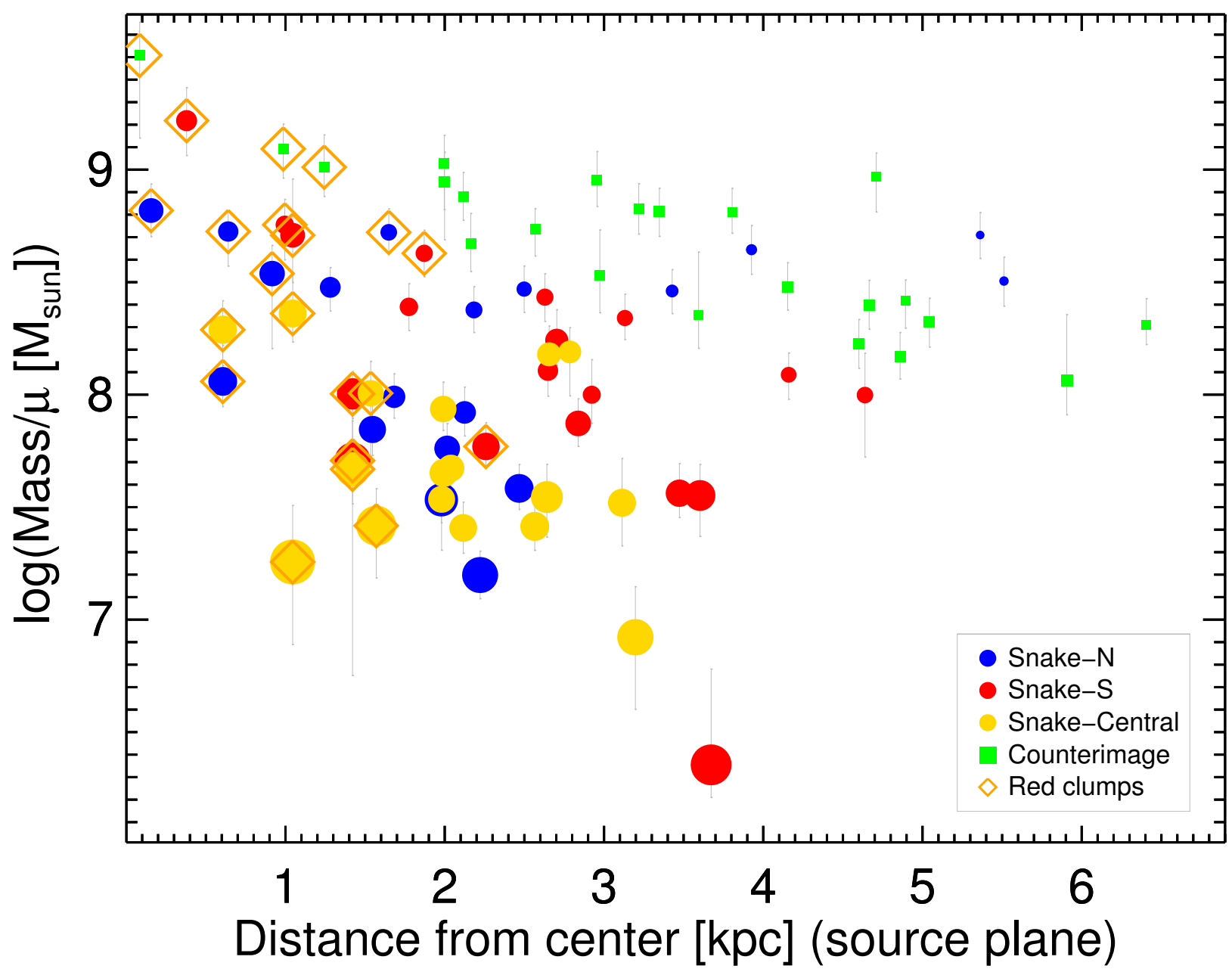

**Figure 3 | Radial distribution of clumps mass in the source plane.** Mass versus galactocentric distance in the source plane of the galaxy for individual clumps detected in the Cosmic Snake and the Counterimage. The size of each point is proportional to the lensing amplification. Error bars for the mass estimate are derived from Monte Carlo simulations.

Finally, examining the contribution of the clumps to the integrated galaxy properties, we note that red clumps contribute $\sim 60$–65 % of the total clump mass but only 15–20 % of the total clump star formation rate. This is again in-line with the picture drawn above where blue clumps are less massive and less dense but have higher star formation rates, whereas red clumps are older and more massive, having increased their masses through merger processes and accretion. Overall the blue and red clumps identified in the Counterimage represent $\sim 20$–30 % of the host galaxy mass and 50–60 % of its global star formation rate. However, we note that the overall contribution from clumps in the Counterimage (the only global image of the source galaxy) is still an upper limit to the intrinsic value, due to the low amplification in this region of the image. A rough estimate from the clumps limited to the Cosmic Snake gives smaller values, 15–20 % of the host galaxy mass and 40–50 % of its global star formation rate, assuming the portion of the lensed source galaxy is representative of the integrated properties. Interestingly, these values are closer to the predictions from recent numerical simulations which take into account feedback mechanisms. Altogether, these results imply that clumps may contribute less to bulge formation and evolution than previously thought.[3] The analysis of a larger sample of multiply-imaged clumpy galaxies by means of the proposed approach is expected to constrain the clumps (and galaxies) properties with unprecedented precision. Strong lensing, at present, provides the only viable approach, waiting for the advent of the next generation of extremely large telescopes (such as the European Extremely Large Telescope, E-ELT, the Thirty Meter Telescope, TMT, or the Giant Magellan Telescope, MGT) that promise to boost our resolution and image depth even more.

**Online Content.** Methods, along with any additional Supplementary Information display items and Source Data, are available in the online version of the paper; references unique to these sections appear only in the online paper.

**Aknowledgements**

The work of AC, DS, MD-Z, LM, and VT is supported by the STARFORM Sinergia Project funded by the Swiss National Science Foundation. JR acknowledges support form the European Research Council starting grant 336736-CALENDS. PGP-P acknowledges support for Spanish Government MINECO grants AYA2015-70815-ERC and AYA2015-63650-P. This work has made use of the Rainbow Cosmological Surveys Database, which is operated by the Universidad Complutense de Madrid (UCM), partnered with the University of California Observatories at Santa Cruz (UCO/Lick,UCSC). Based on observations made with the NASA/ESA Hubble Space Telescope, and obtained from the Hubble Legacy Archive, which is a collaboration between the Space Telescope Science Institute (STScI/NASA), the Space Telescope European Coordinating Facility (ST-ECF/ESA) and the Canadian Astronomy Data Centre (CADC/NRC/CSA).

**Author contributions** Data analysis and interpretation: A.C., D.S., J.R., P.P.-G., M.D.-Z., L.M., V.T. SED fitting: D.S., A.C. Photometry: A.C., P.P-G. Lens Modelling: J.R., A.C. Drafting text, figures, and Methods: The bulk of the text was written by A.C. All authors commented on the manuscript at all stages.

**Competing Interests** Reprints and permissions information is available at www.nature.com/reprints. The authors declare no competing financial interests.

**Correspondence** Correspondence and requests for materials should be addressed to A.C. (antonio.cava@unige.ch)

# METHODS

In this section we provide details about all the aspects of our study presented and discussed in the main article, including: the description of the multi-wavelength data used for the analysis, the definition of the clumps and their size measurements, the spectral energy distribution fitting procedure, and the lensing model providing the amplifications and the geometry at the base of the interpretation of the data.

**Multi-wavelength data.** Public data fundamental for this study include the Cluster Lensing And Supernova survey with HST (CLASH) survey,[17] and other public ancillary including Spitzer and Herschel (from the Spitzer and Herschel Lensing Surveys, PI: E. Egami; *27, 28*), for the longer wavelengths. We also gathered a large amount of additional proprietary data in this field, including SINFONI, PdBI, MUSE and ALMA, which will be analysed in future papers.

The galaxy cluster field is among the targets of the CLASH, which imaged the MACSJ1206.2-084747 cluster in 16 HST bands (4 WFC3/UVIS, 7 ACS, 5 WFC3/IR). The full CLASH dataset and observations are described in *17*. We summarize here the main information regarding the photometric data for the cluster MACSJ1206.2-084747, and show the filter transmission curves in Supplementary Figure 1. Within this field of view we can clearly observe a giant arc (first identified in *14*), that we dubbed as the ”Cosmic Snake” due to its peculiar shape, and its main ”Counterimage” (see Supplementary Figure 2).

We downloaded the fully co-added and mosaicked maps from the CLASH archive,[29] by using the 30 mas version of the maps which ensure for a source with radius comparable to the PSF HWHM to be sampled by at least nine pixels for the finest resolution achievable ($\lesssim$0.1 arcsec) with the shortest wavelength filters.

Exposure time in each HST filter ranges between $\sim$2000s and $\sim$5000s, reaching $5\sigma$ limiting magnitudes (determined randomly placing apertures of 0.4 arcsec) ranging between 26 and 27

AB mag. Note that for smaller apertures the detection limits are deeper, reaching $\lesssim$29 AB mag for $\sim$0.1 arcsec as is the case for several clumps in our study.

The photometry is performed exploiting the *Rainbow Cosmological Surveys Database*[30–33] hosted by the Universidad Complutense of Madrid (see the *Photometry* Section for more details). The Rainbow database is a vast compilation of photometric and spectroscopic data for several of the deepest cosmological fields, such as GOODS-North and South, COSMOS, or the Extended Groth Strip, among others, and tt is publicly accessible through the website.[30]

Data at wavelengths longer than the HST/F160W filter (i.e. Spitzer IRAC/MIPS and Herschel PACS/SPIRE data) are not used in the present study of the clumps properties, since the resolution is too coarse to get any useful information on these scales without performing a full de-convolution which would imply strong assumptions on the SED of the clumps themselves and is beyond the scope of the present work. So, we use those data only for the derivation of the integrated SED of the galaxy, which is well represented by the Counterimage.

**Clumps definition and sizes** In this section we describe the procedure followed to identify the clumps within the magnified galaxy (both in the giant arc and the Counterimage) and we provide details about its structure. The source galaxy is lensed up to five times, including the Cosmic Snake and the Counterimage.

The Cosmic Snake is characterized by a complex shape (a four-folded multiple image) due to the cluster gravitational lensing effect combined to the small-scale deformations induced by the proximity of few cluster galaxies. Despite the much stronger amplification experienced by the Cosmic Snake, since the source galaxy is only partially ($\sim$half) multiply imaged, the geometry of the lensing model (see Supplementary Figure 3 and the *Lensing model* section) results to be very complex in this portion of the image.

The northern and southern portion of the giant arc ("Snake-north" and' "Snake-south" in Supplementary Figure 3, or "Sn" and "Ss" respectively), show the best compromise between high amplification and visible fraction (larger than 50% of the source plane image, see also Supplementary Figure 9) of the projected source galaxy. The two intermediate multiple images (Central-north and Central-South in Supplementary Figure 3, or Cn and Cn respectively) are even more amplified but they show just a very small portion (less than 20%) of the source galaxy. In order to get a global view of the source galaxy we resort to the use of the Counterimage of the galaxy to define the clumps. This global view of the source galaxy is available at the expenses of a reduced lensing amplification (see *Lensing model* Section for details on the lensing model). In fact the average amplification factor for the Counterimage is about 4.5, while for the Cosmic Snake it ranges between $\sim 10$ up to $\gtrsim 100$.

Notably, as mentioned in the main paper, this configuration provides a unique opportunity to observe one galaxy with different physical resolutions set by the different amplifications experienced by the Cosmic Snake and the Counterimage. Such an ideal situation has been considered so far only in galaxy simulations,[4,12,13,16,19] or can be mimicked from the observational point of view by degrading observed galaxies at different resolution through PSF convolution.[34] However, this convolution procedure relies on technical assumptions on the different PSFs and other possible instrumental effects, while we can exploit a *natural convolution* which takes into account all the possible effects at once. Furthermore, the lensing amplification, by preserving the surface brightness, not only provides a better resolution but also a deeper view of the galaxy.

Clump identification in imaging of high redshift galaxies is a complex task. We here consider three different methods, visual selection, isophotal selection, and an 2D-fit method. Each of these methods presents advantages and drawbacks, and the final choice of the method may also depend on the specific characteristics of the galaxies and the objectives of the analysis. Using automated methods (such as isophotal selection, and an 2D-fit method) can result in a more *objective* definition, in principle. However, for individual objects with such peculiar characteristics as the Cosmic Snake, a classical visual definition allows to exploit some advantages that derive from a major flexibility of the method. For example, in our case when selecting clumps we can simultaneously take into account the constraints provided by the lensing model by comparing the predicted positions and colors of the clumps (see details in the Advanced Lensing Modeling section). The fact that several clumps can be cross-identified in multiple images is an additional guarantee and help to the visual identification. Another difference is that for close pairs/groups of clumps we can always visually define a region enclosing each peak, in a way that is mimicked by some automatic method. Finally, visually working on the selection by additionally taking into account isophotal levels allow us to define regions that include all/most of the flux down to a certain level (typically 3-sigma) and at the same time verify the shape and position of the selected clumps. Summing-up all these considerations resulted in our choice of adopting the visual definition for the presented analysis. Thus, we start by discussing this selection method first, and then compare to the isophote and 2D-fit methods to asses the robustness of the selection in the following section. However we stress the fact that in other situations, in particular in the case of large samples of clumpy galaxies, an automatic method would be certainly advisable. The qualitative and quantitative (in terms of number, shape and position of clumps) agreement between the visual selection of clumps and the selection based on isophotes or, when available, measurements from 2D-gaussian fit further support our choice in this specific case.

By visually inspecting the images and selecting the emission peaks along the Cosmic Snake and the Counterimage, we have identified a total of 79 clumps, 19 of which in each the two main portions of the Cosmic Snake (north and south) and 24 in the *Counterimage*. The remaining 17 clumps are distributed in the two central portion of the Cosmic Snake (10 in the Central-north and 7 in the Central-south). The different regions of the Cosmic Snake and the Counterimage are represented in Supplementary Figure 3. We note that the two main portions of the *Cosmic Snake*, Snake-north and Snake-south, just contain about half of the source galaxy due to the complex lensing configuration (see "Lensing model" Section), while in the Counterimage we can observe the full galaxy amplified for a smaller amount. The differential amplification factors in the Cosmic Snake and Counterimage actually allow us in some case to resolve larger star forming regions into sub-clumps, fully exploiting the amplification power provided by the gravitational lensing.

We start from a visual selection procedure based on the CLASH image in the filter F390W. The F390W image is chosen because it represents the best compromise between the resolution, the depth of the imaging, and the contrast between the clumps and the background. In addition this filter represents well the near-UV emission that should trace star forming regions in the galaxy. Of course, any visual approach can introduce subjective biases, but as we will demonstrate in the following by comparing with others methods commonly used in literature (namely the isophote and the 2D-fit - or core - methods, described below), we are confident that the selection procedure does not introduce any large bias both in the selection and the sizes/shapes of the clumps. On the other hand, visually selecting the clumps allows us to optimize the selection and maximize the number of selected clumps.

The 58 visually selected clumps from the F390W filter image are shown as blue ellipses in Supplementary Figure 4 on top of the F390 image. Size and orientation of the elliptical regions are defined to maximize the total flux above $3\sigma$ (per pixel) around each peak, while avoiding the overlap of neighbouring regions. We refer to these regions as *blue clumps*. Based on the minor and major semi-axes, $a$ and $b$ respectively, of the elliptical regions we define an *equivalent radius*, $R_{vis}$, for each clump as: $R_{vis} = \sqrt{a \cdot b}$. These equivalent radii, representing the size of the clumps, are used in all the following analysis.

It is known that different observational bands can probe different physical regions, so we may expect to get different clump definitions at varying wavelengths. In order to obtain (at least partially) a more comprehensive view of the clumps in the galaxy, we also define *red clumps* based on a procedure similar to that followed for the F390W filter but this time applied to the *redder* F110W filter. Supplementary Figure 5 shows the 21 selected clumps as red ellipses on top of the F110W image. Note that, even if the selection is not mutually exclusive, only one red clump would be clearly selected also as blue clump (close to the center of the CI).

The difference in the FWHM ($\sim 0.07''$ for the F390W versus $\sim 0.13''$ for the F110W filter) between the two selection filters can in principle affect the size definition of the clumps region. On the other hand, since we want to extract the photometry from PSF-matched images, we assume the FWHM of the worst PSF (FWHM$_{F160W}$ $\sim 0.16''$) as a lower limit on the apertures (diameter) definition. This assumption ensures that we probe equivalent physical regions in all the bands and that we do not under-sample the PSF in all the cases providing a PSF-correction at maximum of $\sim 10\%$ for the sizes. Ideally, it would be advisable to perform the clumps selection directly in the PSF-matched images, but doing that would not allow selecting all the clumps seen in the native (higher resolution) images due to the smearing of the PSF. The minimum aperture size limit allows to homogeneously couple the selection to the photometry. However, we note that our results would hold completely unaffected (or would be even strengthened) by reducing the sizes of the few clumps (only three) affected by this limitation, since all the relevant conclusions of the analysis are constrained by the upper limits of the clumps sizes and masses (as discussed in the main paper). In order to qualitatively check the robustness of this visual clumps definition, we contrast the clumps selected in the F390W and F110W filters with the co-added ACS+WFC3 image which provides higher S/N data, at the expenses of degrading the resolution (due to the wider PSF for longer wavelengths). We visually confirm the identification of each clump in the co-added image, despite the slightly lower resolution due to the smearing of the PSF when co-adding different HST bands.

We rely on the sizes and shapes visually defined from the two selection filters and, when possible, compare the results to other approaches (isophote and 2D-fit methods). We note that, in the following, we will use the term *size* as equivalent to the *circularized radius*, $R_{vis}$, defined above in this Section.

**Comparison of different clump definition approaches** In the literature, the size of the clumps has often been defined from the area above a given surface brightness level. This isophote method has a few important drawbacks. First of all, the surface brightness threshold is typically determined visually, from a trade-off between identifying a maximum number of regions while minimizing blending between individual regions. The chosen isophote is thus subjective and difficult to compare between studies, especially over a range of redshifts. Secondly, this method can be influenced significantly by local background variations, especially at high redshifts where undetected low surface brightness regions or overlapping light from regions that are only separated by a few pixels can enhance the local background level. However, this method is non-parametric and provides a good solution for bright and/or nearby objects where structures are typically better defined and separated. An independent robust way to measure region sizes is provided by the core method, in which a 2D light profile is fitted to the surface brightness profile of each region (see *15* for a description and comparison of the two methods). The local background is a free parameter in the profile fit, thus minimizing its influence on the size measurement. Most commonly, a 2D Gaussian light profile is used as parametrization, which is a good approximation to probe the central ionized core of the HII regions. However, when working on galaxies with a high diffuse background level and with clumps spatially close to each other this fitting method starts to encounter more difficulties in the convergence of the fits. This is in particular the case for our Cosmic Snake and Counterimage where many neighbouring clumps are identified deeply embedded in the diffuse galaxy background. For these reasons, both the isophote and 2D-fit methods are not ideal in our case, since the number of clumps, their small sizes, their proximity and the high background noise and flux levels make difficult to select a single reasonable limit for the isophote or prevent the convergence of the fitting ap-proach in most cases. Visually defining elliptical shapes for the regions allows to combine the advantages of other two methods, in a sort of parametric isophote approach.

As a quantitive check on the shape and size definition of the clump regions, we perform a gaussian 2D fit using the publicly available code *iGalfit*.[35,36] This code relies on GALFIT[37] (version 3.0) to create a model for each individual star-forming region in the HST image. In Supplementary Figure 6, we compare the clump sizes obtained from the visual definition and the 2D-gaussian fit, $R_{iGal}$ (defined from the minor, x, and major, y, axes HWHMs of the 2D-gaussian fit similarly to the visual equivalent radius as $R_{iGal} = \sqrt{HWHM_x * HWHM_y}$), for the regions for which the fit converge ($\sim 30\%$ of the sample). As anticipated, in several cases the fit does not converge due to the proximity of different clumps which are difficult to separate with an automatic method, or due to the high background level. We note that, it is also possible that the fit does not converge because some of the clumps identified by eye are not real clumps. However, such convergence problems using automatic methods in high-background images are among the known drawbacks of these methods. Of course, a much more quantitative analysis could be performed using simulations and/or adding sources to the HST images, but this is beyond the goals of this work.

For the clumps for which we can get a robust 2D fit estimate from *iGalfit* ($\sim 30\%$ of the sample), we see that the agreement between visual-based and fit-based sizes is fairly good. We get a robust linear correlation between to two definition of radii, suggesting just a possible small bias ($\lesssim 5\%$) and a $\sim 20\%$ scatter in the ratio of the radii (which we can assume as typical, conservative, error on the size estimates). The small bias, indicating slightly larger sizes determined by visual inspection is expected given the fact that the visual selection of clumps tends to probably include more background flux, similarly to the isophote method.

As a last check, we also compare our visual definition with an isophote definition, assuming a $\sim 3\sigma$ cut on the image (see white contours in Supplementary Figure 4 and 5). We find an overall good agreement between the two selections ($\sim$85% of the sample overlap), with the visual selection able to select more clumps mostly based on the support from the lensing model (symmetries and predictions in the positions with respect to the critical lines, see Advanced lensing modeling Section for the details). From the visual selection we obtain 21/14/12/7/4 blue and 3/5/7/3/3 red clumps for the Counterimage and Snake-N/Snake-S/Central-N/Central-S portions of the Cosmic Snake respectively. Of the 21/14/12/7/4 blue and 3/5/7/3/3 red visually identified clumps, 18/12/11/6/4 blue and 3/4/5/2/2 red clumps are also identified using the isophote method (applying a 3-sigma cut). That is, the visual selection detects 3/2/1/1/0 blue and 0/1/2/1/1 red independent additional clumps. On the other hand, the isophote selection would select 4/2/3/1/1 additional blue clumps (in the the Counterimage/Snake-N/Snake-S/Central-N/Central-S respectively) and only 1 additional red clump in the Counterimage (none in the Snake) with respect to the visual selection. All the clumps additionally selected by the isophote method, and not by visual selection, are extremely faint (just above the 3-sigma isophote selection cut) and small (few pixels). These clumps could likely be noise peaks in the images. Only two of these clumps would be also supported by the lensing model inspection (see Advanced lensing modeling Section for details), however we preferred to keep only visually selected clumps in the final catalog in order to adopt a unique selection method. On the other hand all the clumps additionally detected by the visual selection originate from the split of extended isophotes or clumps at the limit of the detection which are also supported by the symmetries and predictions in the lensing model (see Advanced lensing modeling section for details). Globally 85% of the clumps visually selected are confirmed as independent peaks above 4-5 sigma using the isophote selection. The remaining cases ($\sim$15% of the sample) would have blended or low significance isophotes having 3 to 4 sigma peak emission. The visual selection, coupled to the lensing model, is able to de-blend and confirm these cases with good accuracy, as independently confirmed by the direct comparison of cross-matched clumps sizes given in the lensing modeling section.

Most of the clumps are marginally, or well, resolved ($\sim$50% have diameters $\geqslant 1.35$ times the $PSF_{\rm F160W}$), as can be seen from Supplementary Figure 7 where we plot the relative resolution (defined as the ratio between the clump size and the PSF size for the HST-F160W filter) as a function of the clump radius expressed in physical units (top panel) and clump mass (bottom panel) after correcting for lensing magnification. Independently from the magnification, there are not evident biases with respect to size, the mass, or the sub-sample (blue/red) of clumps. Supplementary Figure 7 provides an estimate of the intrinsic level of resolution that we can achieve for each clump, showing that for the coarsest resolution band (F160W) about half of the sample would be still unresolved. The situation improves moving to shorter wavelength bands, where the resolution progressively increases and the PSF FWHM attains a minimum value of $\sim 0.07''$in the best cases. For these higher resolution filters, in principle all the individual clumps would be resolved, but we note that it would not be possible to distinguish between a genuinely individual resolved clump and a small cluster of unresolved clumps. For this reason, when talking about clump sizes (and as a consequence for the masses) we should keep in mind that from a physical point of view we might still be looking at upper limits, due to the finite value of our instrumental resolution.

In Supplementary Table 3 we report the clumps properties, including those obtained from the SED fitting and lensing model described in the following Sections: ID, semiminor and semi-major axes (in arcsec), physical size (in terms of circularized radius $R_{vis}$, in pc after applying the magnification factor $\mu$), mass, and magnification factor.

**Photometric extraction.** Using the clumps selection and definition described in the previous Section, we extract aperture photometry for each region exploiting the *Rainbow* tools.[31] The first step is to PSF-match all the 16 HST observed filters to a common PSF, which we assume to be equivalent to the coarsest resolution provided by the WFC3-F160W filter ($\sim$ 0.16”).

Photometric extraction is then performed on this PSF-matched images, in order to guarantee the sampling of the same physical regions in all the bands. Note that the CLASH images are aligned so that the objects are sampled by the same pixels for all the filters.

Aperture photometry was performed with the Rainbow Database tools described in Pérez-González et al. (31; see also 32, 33). Briefly, the photometry code uses elliptical apertures, considering fractions of pixels, an important feature when dealing with small apertures such as the ones for the Cosmic Snake clumps. The background (median value and rms) is estimated in a rectangular area around the source, $1'$ on a side. In order to deal with the correlated noise introduced by the drizzling reduction method, the background is estimated by building artificial apertures of the same size and shape as the considered clump, using randomly selected pixels around the source. Finally, aperture corrections are computed for each region by measuring the PSF's encircled energy fraction at increasing radii, from $0.02''$ up to $2''$ in small steps of $0.02''$, and applying the corresponding correction to the aperture size. The average correction applied to the measured photometry result to be $<\mathrm{AP}_{corr}> \sim 2.5$, with an rms $\sim 0.6$. The photometry is then corrected for Galactic extinction[38] before using it for the SED fitting.

The measured photometry is naturally affected by the application (or not) of a local background, which necessarily is higher than our "sky" background. However, in PSF-matched images the local background is dominated by the wings of the PSF due to the fact that typically clumps are marginally resolved and the size of the apertures is comparable to the PSF-FWHM (i.e. local background is dominated by the clump light scattered by the PSF itself). Therefore, even if the fluxes are intrinsically lower the colors must remain unchanged. This fact implies that, since we have not applied a local background subtraction to the photometry of the clumps, parameters such as the stellar mass of the clumps might be overestimated because part of the flux might be coming from foreground and background regions in the same galaxy. This effect could be as large as 20–50% on the measured properties, according to *5, 22*, and highly depen-dent on whether the clump is near the 'bulge' of the galaxy or at large galactocentric distance. Assuming the worst case, 50% contribution due to the background, we derive that the maximum variation in the mass estimates for clumps in the central region of the galaxy is about 0.3 dex. This contribution is progressively reduced moving from the center toward the external regions of the galaxy. The cumulative effect will be a shallower slope (after background subtraction) of the mass-radial distance trend. Since the variation between the maximum (galaxy center) and minimum (external regions) radial mass distribution (as observed in Figure 3) are larger than 1 dex (in the worst case, i.e. for the Counterimage), we can deduce that background variations and subtraction does not significantly affect our conclusions. In other words, the observed trend cannot be mimicked and explained by the contribution of the background, at most the observed trend could be slightly shallower, but we are not interested here in the absolute slope. Finally, our conclusion on the maximum clump mass would only be reinforced if the background contribution was not negligible.

We list the clumps and integrated Counterimage photometry in Supplementary Table 1 and 2.

**Spectral energy distribution fitting and determination of physical parameters.** We adopted a modified version of the code Hyperz[18,39] to perform the SED fit of the photometric data extracted for each clump. For the clumps we fit 15 of the 16 bands of the HST photometry, leaving out the bluest filter (F225W), where none of the clumps is detected. To determine the total stellar mass of the galaxy we use again the 15 HST bands plus the IRAC photometry at 3.6 and 4.5 $\mu$m.

The spectroscopic redshift of the galaxy is known, i.e. fixed to the spectroscopic value (z= $1.036$; *14*) in the fits. We have adopted Bruzual&Charlot[40] stellar tracks at solar metallicity. The metallicity, assumed from the mass-metallicity relation at $0.9 \lesssim z \lesssim 1.2$ from Pérez-Montero et al.,[41] is $12 + \log(O/H) \simeq 8.8$, marginally sub-solar (solar value ~8.9). This is also confirmed by direct measurement of the N2 index[42] for the Cosmic Snake using SINFONI data, which results to be $\log$([NII]/H$\alpha$)=0.35 $\pm$ 0.09, with small variation ($\sim$20–30% ) along the arc. We adopt the Salpeter[43] initial mass function over the mass range of 0.1 to 100 $M_{\odot}$. We have allowed for variable star formation histories, parametrised by exponentially declining models with timescales varying from 10 Myr to infinity, corresponding to a constant star formation rate. More precisely, we have used the following timescales $\tau = 0.01$, 0.03, 0.05, 0.07, 0.1, 0.3, 0.5, 0.7, 1., 3., $\infty$ Gyr. This parametrisation is commonly used in the literature, including in our recent homogeneous analysis of clump masses,[10] and allows thus meaningful comparisons with earlier results. The age is a free parameter in the fits, imposing no minimum age. Nebular emission has been neglected in our default models, as it was found not to affect the resulting stellar masses. The attenuation is described by the Calzetti[44] law and is varied from $A_V = 0$ to a maximum value $A_V^{\rm max} = 2$ in steps of 0.05 mag.

Stellar masses are well constrained by the multi-band photometry extending up to $\sim 8000$ Å restframe. Monte Carlo simulations show typical uncertainties of 0.1 to 0.3 dex maximum for the individual clumps. Star formation rate (SFR) and age (defined as the time since the onset of star formation) are more uncertain. As well known[45] they depend strongly on the assumed star formation history and are prone to some degeneracy between age and dust attenuation. The maximum attenuation allowed for the clumps, $A_V^{\rm max}$, is in fact constrained by the observed global IR luminosity, $\log(L_{\rm IR-\mathit{OBS}}) = 12.1 \pm 0.1$ $L_{\odot}$. This value has been determined fitting a modified blackbody to the IR data from Herschel (PACS 110 $\mu$m and 160 $\mu$m) and Spitzer (MIPS 24$\mu$m) and integrating the solution between 8 $\mu$m and 1000 $\mu$m (rest-frame wavelengths). The observed IR luminosity is in agreement, within 1-$\sigma$ uncertainties, also with that predicted from the optical SED fit ($\log(L_{\rm IR-\mathit{SED}}) = 12.0 \pm 0.3$ $L_{\odot}$) and based on template fitting. It turns out that allowing for $A_V^{\rm max} > 2$ produces SED fits with very young age ($\ll 10$ Myr) and hence a large UV attenuation for some clusters. In this case their predicted IR lumi-nosity (from re-emission of the absorbed UV radiation) exceeds the total IR luminosity of the entire galaxy. This justifies our adopted value of $A_V^{\max} = 2$.

**Advanced lensing modeling.** The cluster MACSJ1206.2-084747 has been extensively studied in literature with both dynamical[46] and lensing[47] modeling. We performed an optimization of a model for the total cluster mass distribution constrained with strong lensing, starting from a previous version published in Ebeling et al.[14] We use 6 groups of clumps matched from their color and morphology in the top and bottom region of the Cosmic Snake (totalling 16 lensed images) as individual strong lensing constraints, as well as another triple system in the eastern side of the cluster (systems 2 and 3 in *48*), which is spectroscopically confirmed at $z = 3.038$ (see also *49*). The list of individual constraints used in the model is presented in Supplementary Table 4.

Our mass model is parametric and accounts for both cluster-scale and galaxy-scale mass contributions, which are modelled as multiple pseudo-isotermal elliptical potentials. Individual cluster members are selected from their HST photometry through a red sequence selection, and assigned individual mass components following the elliptical shape (center, ellipticity and position angle) of their light distribution as measured by SExtractor. Their individual central velocity dispersions $\sigma_0$ and truncation radii $r_{\rm cut}$ are scaled according to their luminosity with respect to an $L^*$ galaxy at the cluster redshift assuming a constant mass-to-light ratio. In addition, the model accounts for smooth large scale mass distributions representing the contribution from the cluster dark matter halos: for such potentials we optimise the geometrical parameters (position and shape) as well as the central velocity dispersion $\sigma_0$ and a core radius $r_{\rm core}$. Their large cut radius is unconstrained and we fix it to 800 kpc: changing this to any large value has no effect on the best fit parameters of the model.

Four cluster members located at the vicinity of the Cosmic Snake have a direct influence on the predicted locations of the clump positions and their amplification. They are marked by crosses in Supplementary Figure 8. We individually optimised the main mass parameters of their associated potentials (ellipticity, position angle, $\sigma_0$ and $r_{\rm cut}$) instead of assuming the constant M/L relation, in order to improve the match with the observed centers and shapes of the clumps. Similarly, we individually optimised $\sigma_0$ and $r_{\rm cut}$ for the brightest cluster galaxy (BCG), which typically does not follow the same M/L relation as the rest of cluster members.[50]

We make use of Lenstool[51] to optimise the model parameters according to the strong lensing constraints. Lenstool makes use of a Monte-Carlo Markov Chain (MCMC) bayesian optimisation to explore the parameter space. The goodness of the model is quantified from the rms between the predicted and observed locations of the strong lensing constraints as measured in the image plane.

We first model the smooth mass distributions with a single cluster-scale elliptical component as in Ebeling et al.[14] This symmetric model did not perfectly reproduce both the Cosmic snake and the multiple system to the east (2.1, 2.2, 2.3), with an rms of $\sim 1.2$" Indeed, the distribution of cluster members is asymmetric with respect to the BCG and extends to the East. We then include the contribution of a second elliptical halo centered around the second brightest BCG to the east, greatly improving the rms to $0.15$". All the best fit parameters of the mass distribution are presented in Supplementary Table 5.

As additional checks on the reliability of the model, we predict the shape of the critical line at $z = 1.036$, which is dependent on the local perturbations by individual cluster members. It perfectly matches the symmetries seen in the locations of the clumps (Supplementary Figure 8). We also reconstruct the source plane morphology of the Cosmic Snake based on the top and bottom part of the long arc as well as the counterimage to the north finding a very good match between the different reconstructions (Supplementary Figure 9). Furthermore, as a more quantitative test, we compared the sizes of 18 couples of clumps with robust visual cross-identification between different parts of the image. We find a remarkable agreement be-tween the de-magnified sizes of these clumps, with an almost one-to-one relation (Spearman's test gives p-value $\sim 1.7$e-7) with $\sim$30% scatter, compatible with the inferred typical error on the size estimates (20% relative error, to which we should add the typical uncertainty on the magnification values of the order of $\sim$10%–20%). This check, provides independent support to the robustness of our clump selection and size definition. Finally, we can re-simulate in the image plane the expected pixel distribution of the straight arc in all 16 CLASH/HST bands based on these source plane reconstructions: again we find a very good match between the overall predicted and observed colors and morphologies along the arc, even in between the clumps (Supplementary Figure 10).

We used the advanced lensing modeling presented in this section also to further support and assess the reliability of our visual selection of clumps. After visually defining all the clumps across the Cosmic Snake (as described in the Clumps definition Section), we consider all the clumps defined only in the southern portion of the Cosmic Snake and de-project them to the source plane using our fiducial lensing model. Then we project back these regions from the source plane to the image plane, so that we derive all the predicted positions for the clumps in the image plane. We compare the predicted positions with those of the clumps visually selected in the first step to check for possible counterparts. The same procedure is repeated using the other portions of the Snake to further confirm the clumps identification. Color composite images (such as Supplementary Figure 10) are also used in the cross-identification of clumps to help and support the selection, in fact gravitational lensing is color invariant so that corresponding clumps preserve their colors independently on the specific portion of the Cosmic Snakeunder consideration.

**Data Availability.** The data that support the plots within this paper and other findings of this study are available from the corresponding author upon reasonable request.

**Code Availability.** To perform spectral energy distribution fitting we have used a modified version of the *Hyperz* code, available in its original form at http://webast.ast.obs-mip.fr/hyperz/. For the photometric extraction, we made use of the *Rainbow Database tools*, which are not publicly available as source code, but can be accessed through an interactive interface at https://rainbowx.fis.ucm.es/Rainbow_navigator/.

To perform checks on the size measurements we have used *iGalfit*, publicly available at http://dls.physics.ucdavis.edu/∼rer/iGalFit/igalfit_v1.0/www/home.html.

The lensing model has been obtained using *Lenstool*, publicly available at https://projets.lam.fr/projects/lenstool/wiki

**Additional references**

# Supplementary Information

This section contains all the supplementary Data (Figures and Tables) supporting the analysis presented in the Method section. We include from Supplementary Figure 1 to Supplementary Figure 10 and from Supplementary Table 1 to Supplementary Table 5.

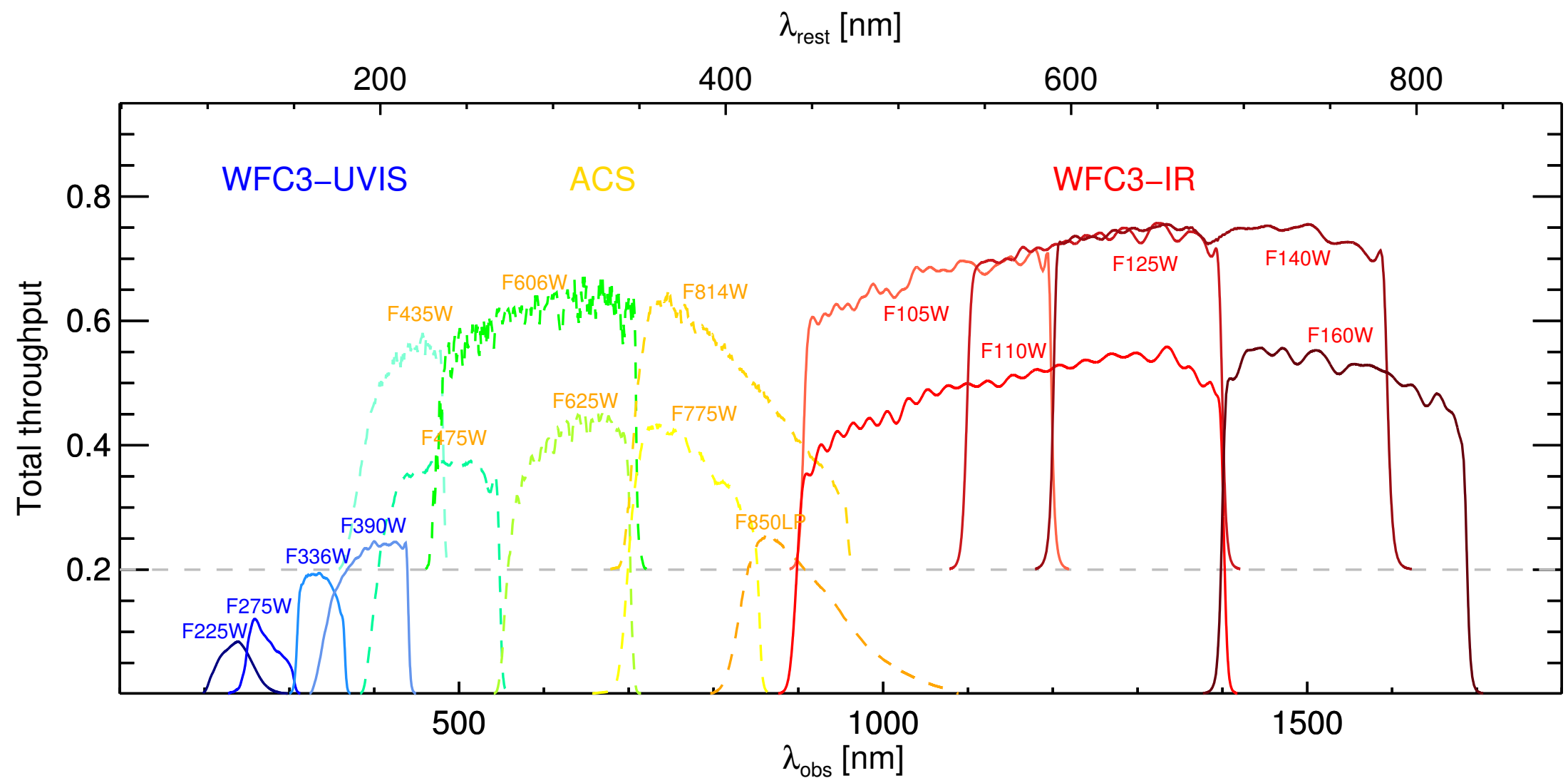

**Supplementary Figure 1** | Filter transmission curves for the 16 HST observed filters. Total throughput curves are plotted for each band, with some of the curves offset vertically by 0.2 (horizontal dashed line). ACS filters are shown as dashed curves, for clarity. Top X-axis gives the rest-frame wavelength at the redshift of the source galaxy (z=1.036).

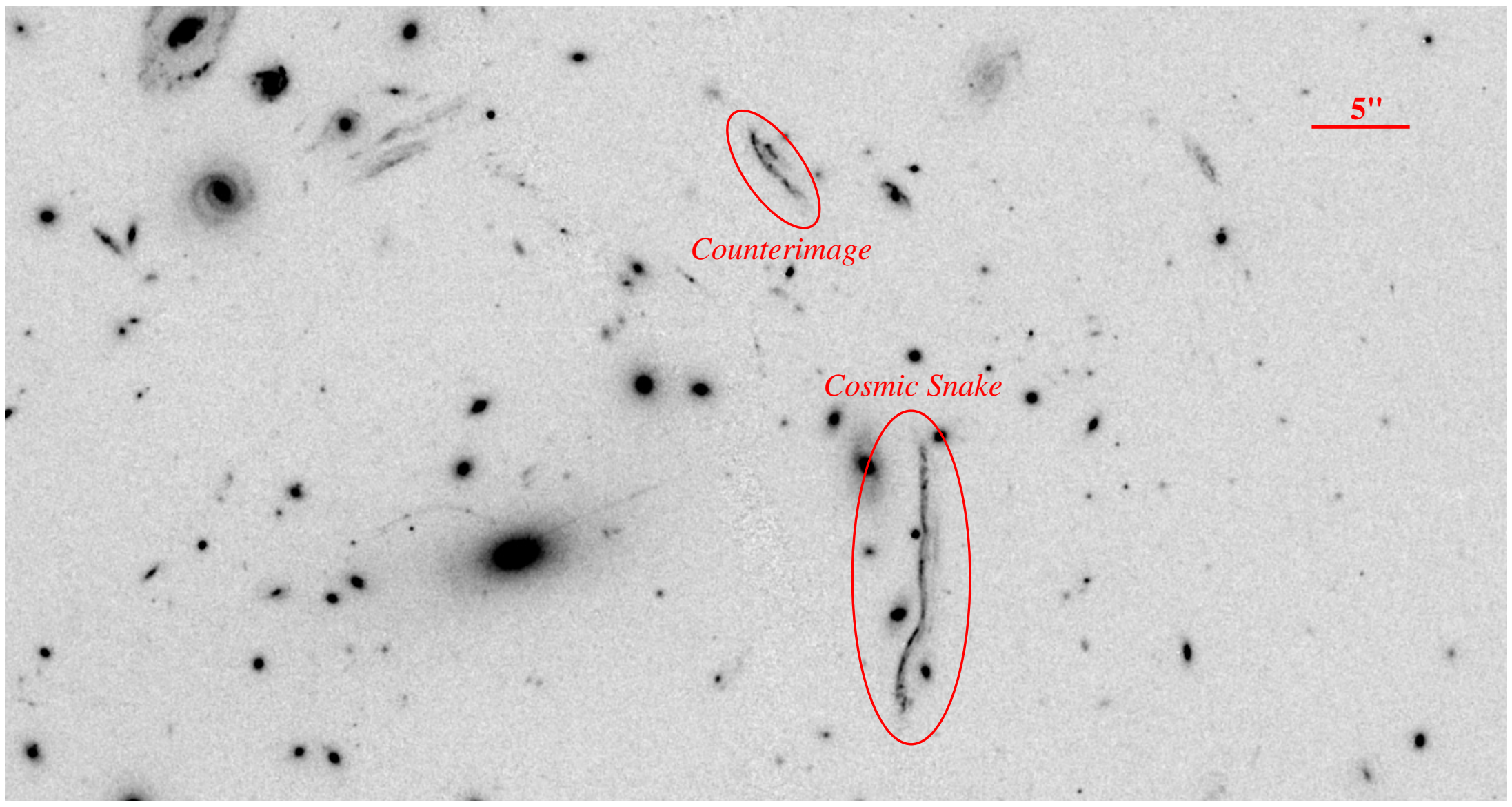

**Supplementary Figure 2** | Portion of MACSJ1206.2-0847 field (F606W) including the giant arc (bottom ellipse) and its counterimage (top ellipse). Image orientation: North-top, East-left. As a reference a scale bar is overplotted on the top-right.

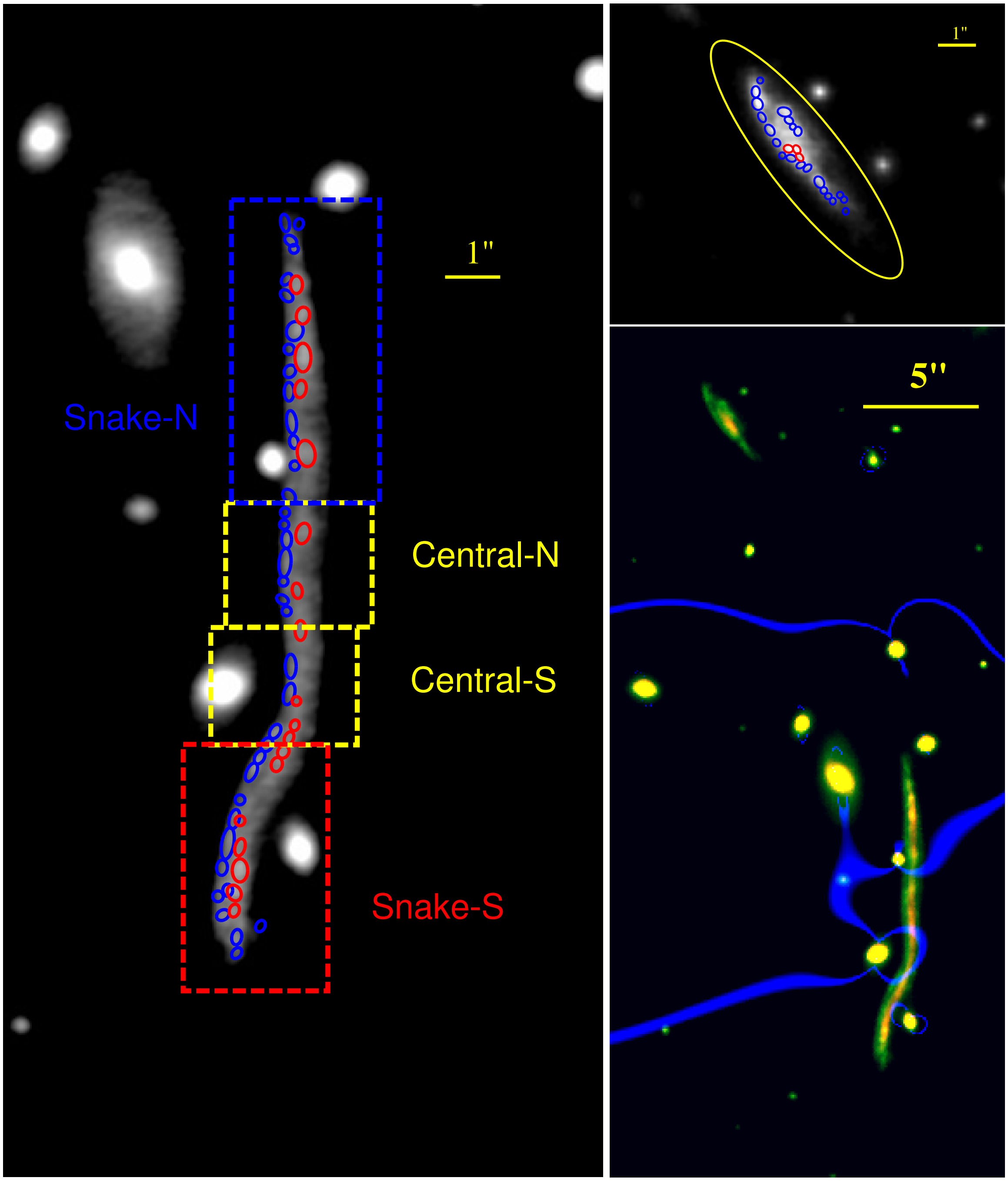

**Supplementary Figure 3** | The Cosmic Snake (left panel) and its Counterimage (top-righ panel) with regions defining clumps: blue regions for blue clumps, red regions for red clumps, yellow region for whole galaxy. Rectangular areas define the four portions of the Cosmic Snake identified from the lensing model. *Bottom-rigth panel:* RGB composite image including: red = F160W, green = F110W, blue = amplification map. For the fiducial lensing model, blue shaded areas indicate amplification above 100, close to the critical lines. Representative scale bars are provided in each panel.

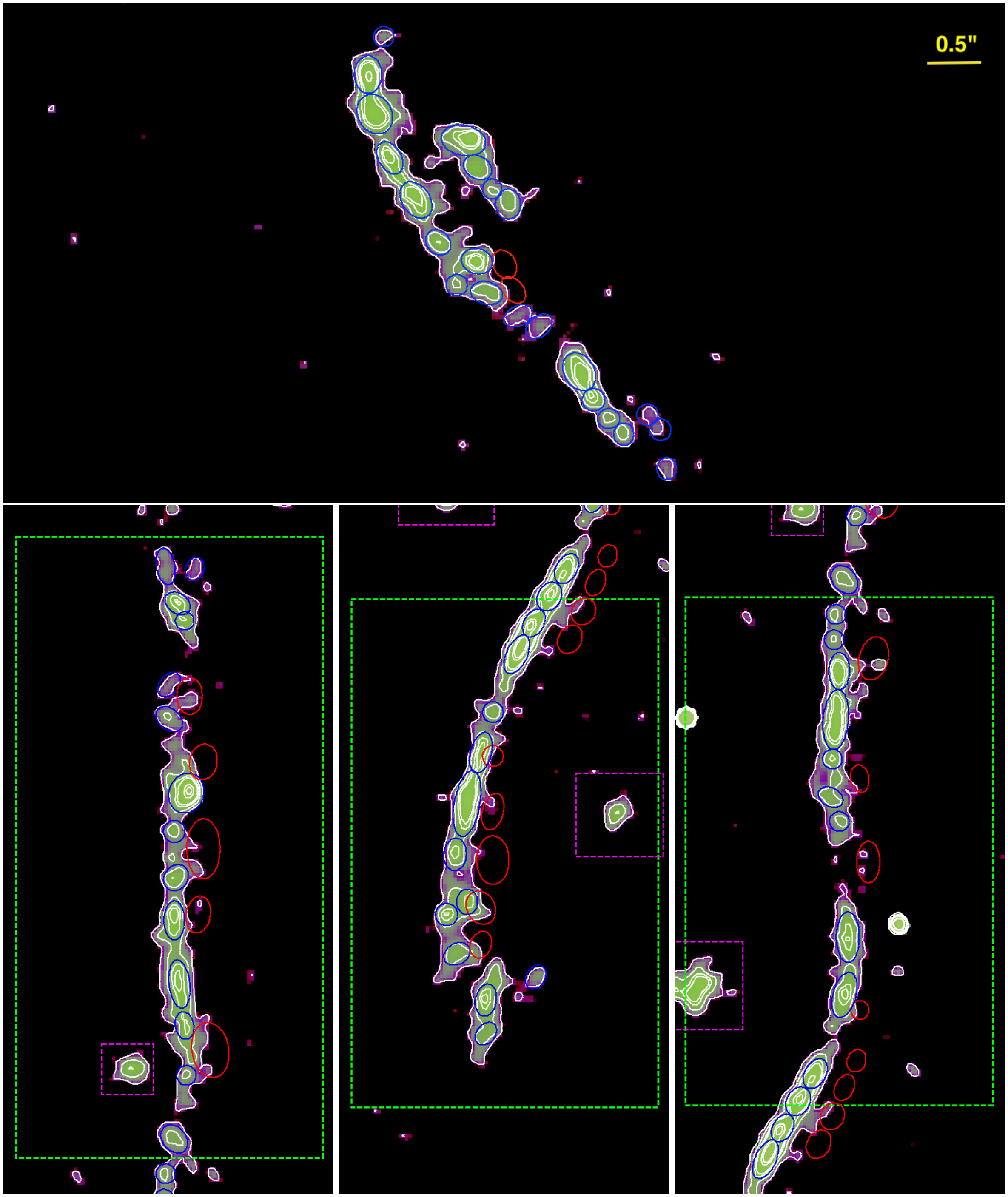

**Supplementary Figure 4** | Filter WFC3-F390W. Zoom on the Counterimage (top panel) and different parts of the Cosmic Snake (bottom panels, North, South and central part from left to right) to highlight the region definition from visual (blue and red ellipses) and isophotal selection (white contours; lowest isophotal level at $\sim 3\sigma$). A reference scale bar is overplotted on the top-right of the top panel.

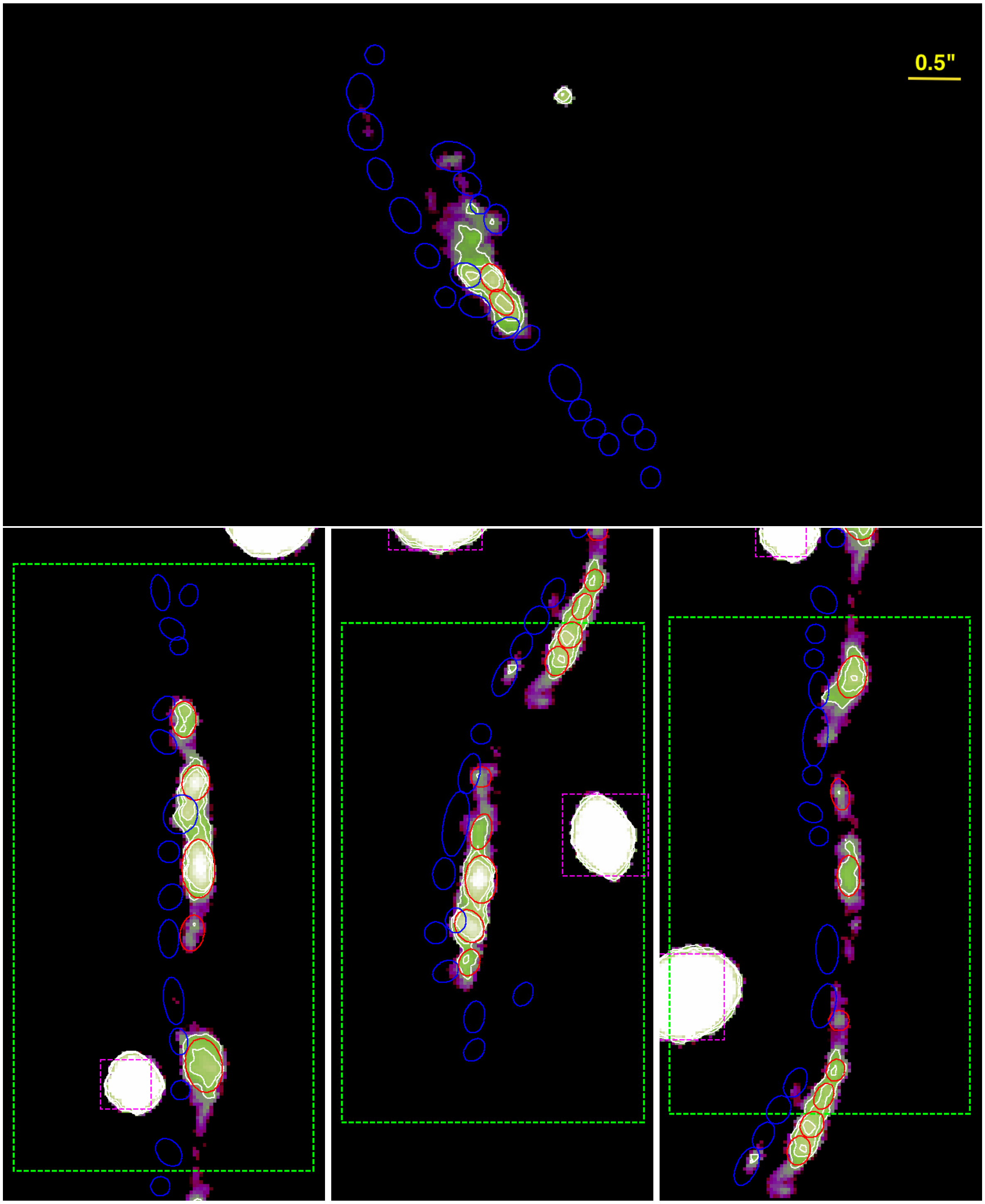

**Supplementary Figure 5** | Filter WFC3-F110W. Zoom on the Counterimage (top panel) and different parts of the Cosmic Snake (bottom panels, North, South and central part from left to right) to highlight the region definition from visual (blue and red ellipses) and isophotal selection (white contours; lowest isophotal level at $\sim 3\sigma$). A reference scale bar is overplotted on the top-right of the top panel.

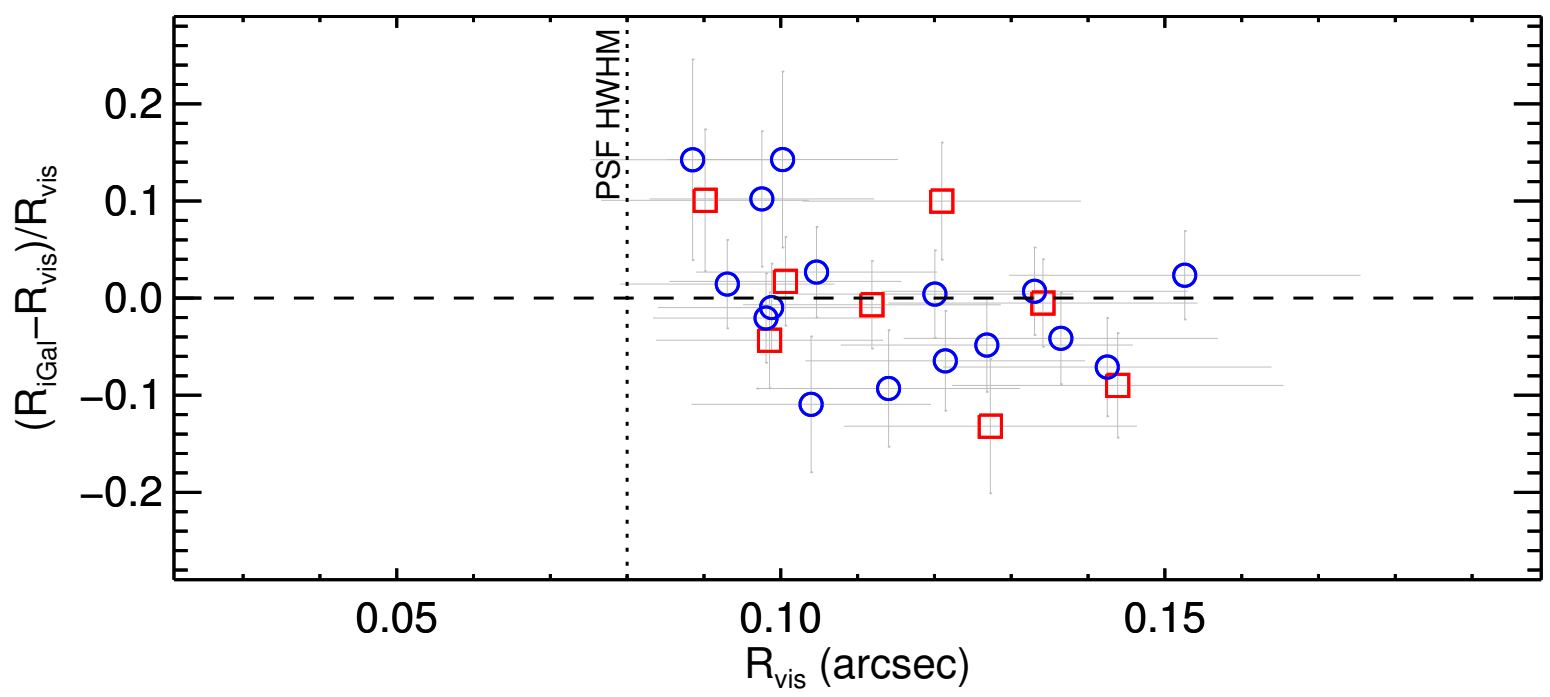

**Supplementary Figure 6** | Comparison of visual estimate for the circularized radius of the clumps ($R_{vis} = \sqrt{a \cdot b}$) and 2D-gaussian fit using *iGalfit* ($R_{iGal} = \sqrt{HWHM_x \cdot HWHM_y}$), for clumps with both estimates. The vertical line shows the lower limit for the clump radius set by the PSF HWHM. Red squares for red clumps, blue circles for blue clumps. Error bars are derived from the relative error budget for the sizes and using standard error propagation.

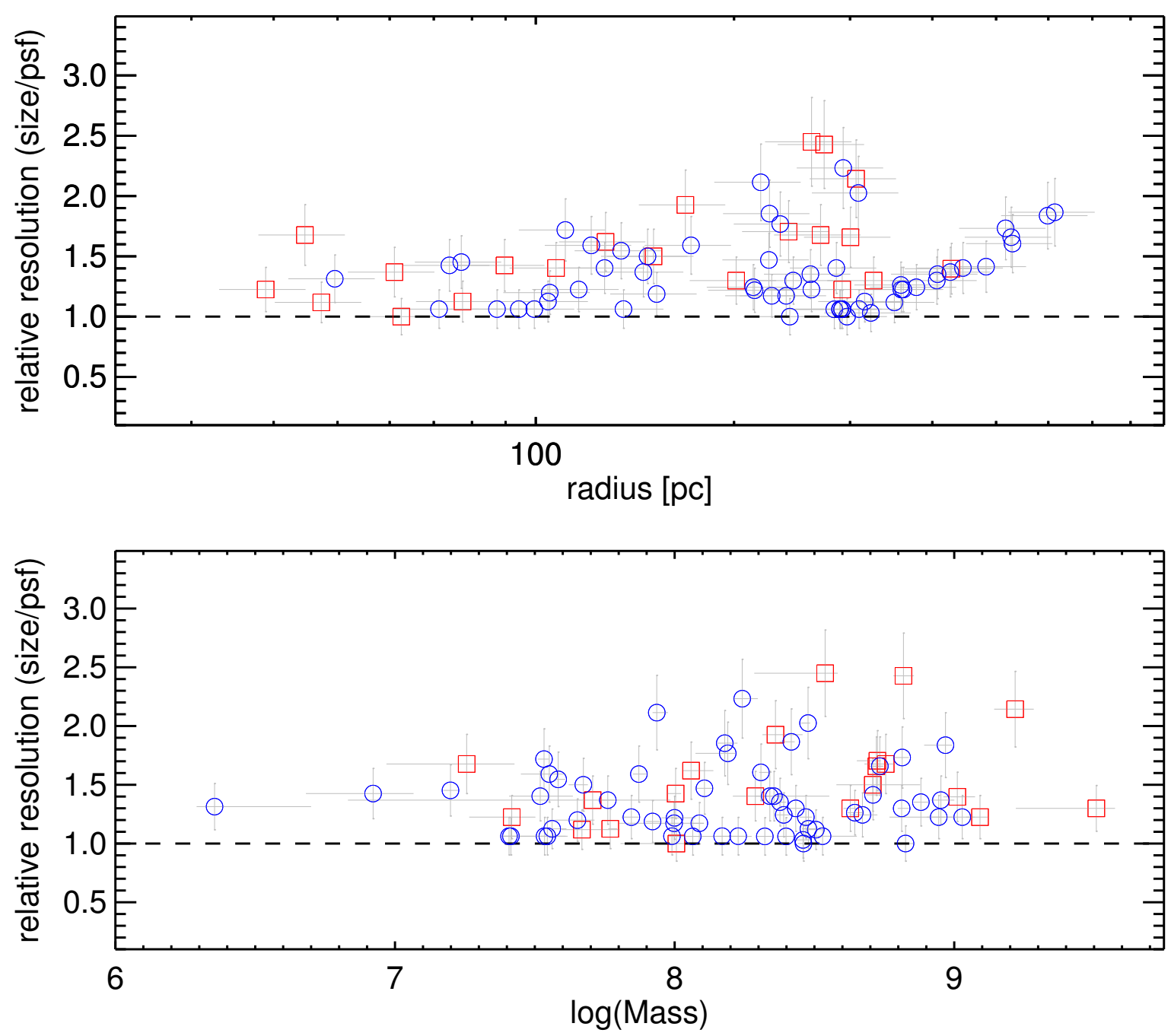

**Supplementary Figure 7** | Relative resolution, defined as the ratio between the size of the clump in the source plane and the de-lensed PSF FWHM as a function of the clump radius in physical scale (top panel) and the (bottom panel) clump mass. The horizontal dashed line sets the limit for resolved clumps, and upper limit for unresolved clumps. Red squares for red clumps, blue circles for blue clumps. Error bars are derived from the relative error budget for the sizes and using standard error propagation. For the mass estimate, error bars are computed from Monte Carlo simulations.

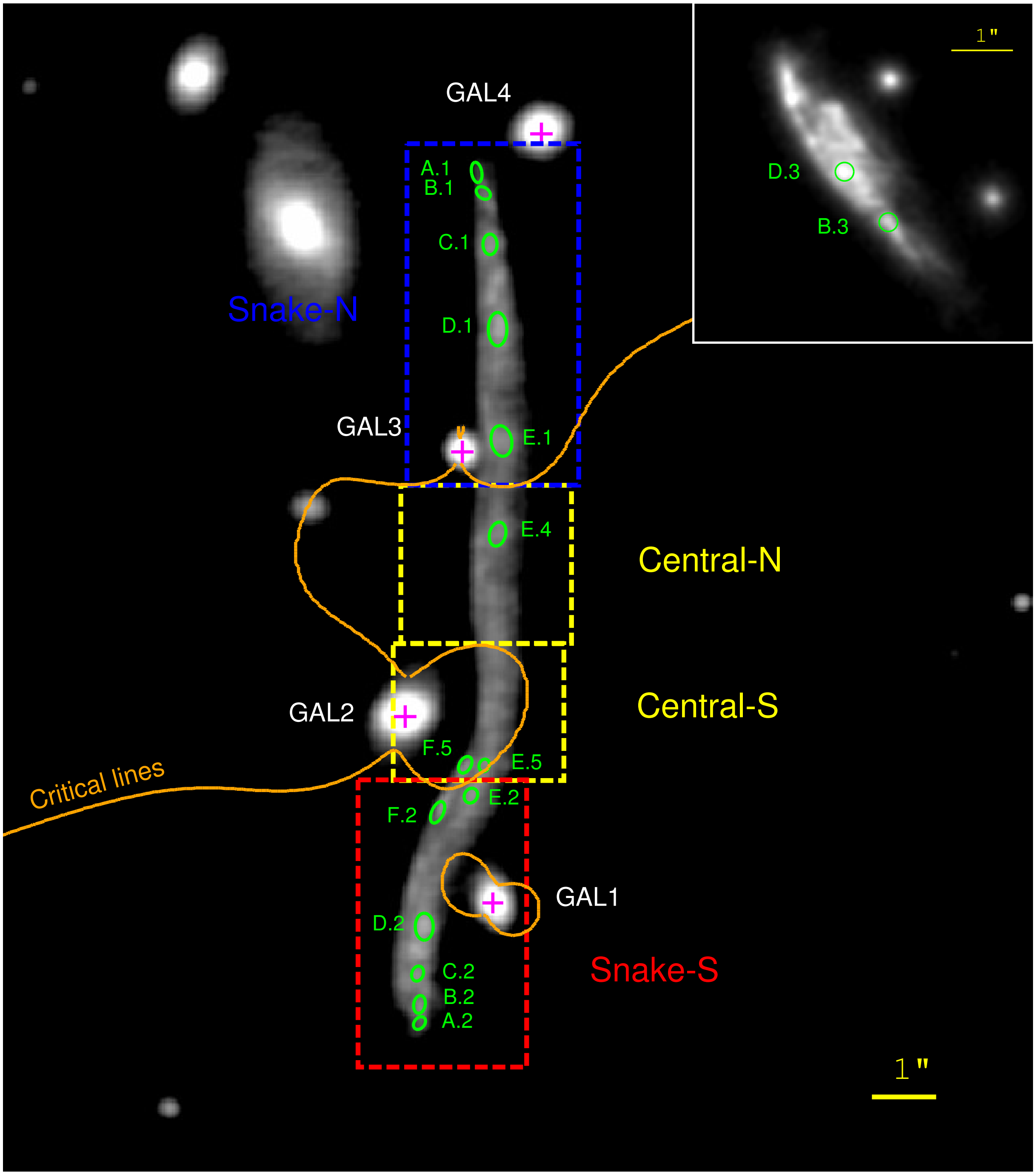

**Supplementary Figure 8** | Detail of the lensing model defining the different parts of the Cosmic Snake (blue, yellow , and red boxes), overlaid to the co-added WFC3+ACS image. We also mark the regions used as constraints for the lensing model (green regions; cf. Supplementary Table S4), the critical lines (orange curves), and the four cluster members mainly affecting the model. The inset shows the constraints for the Counterimage. A reference scale bar is overplotted in each panel.

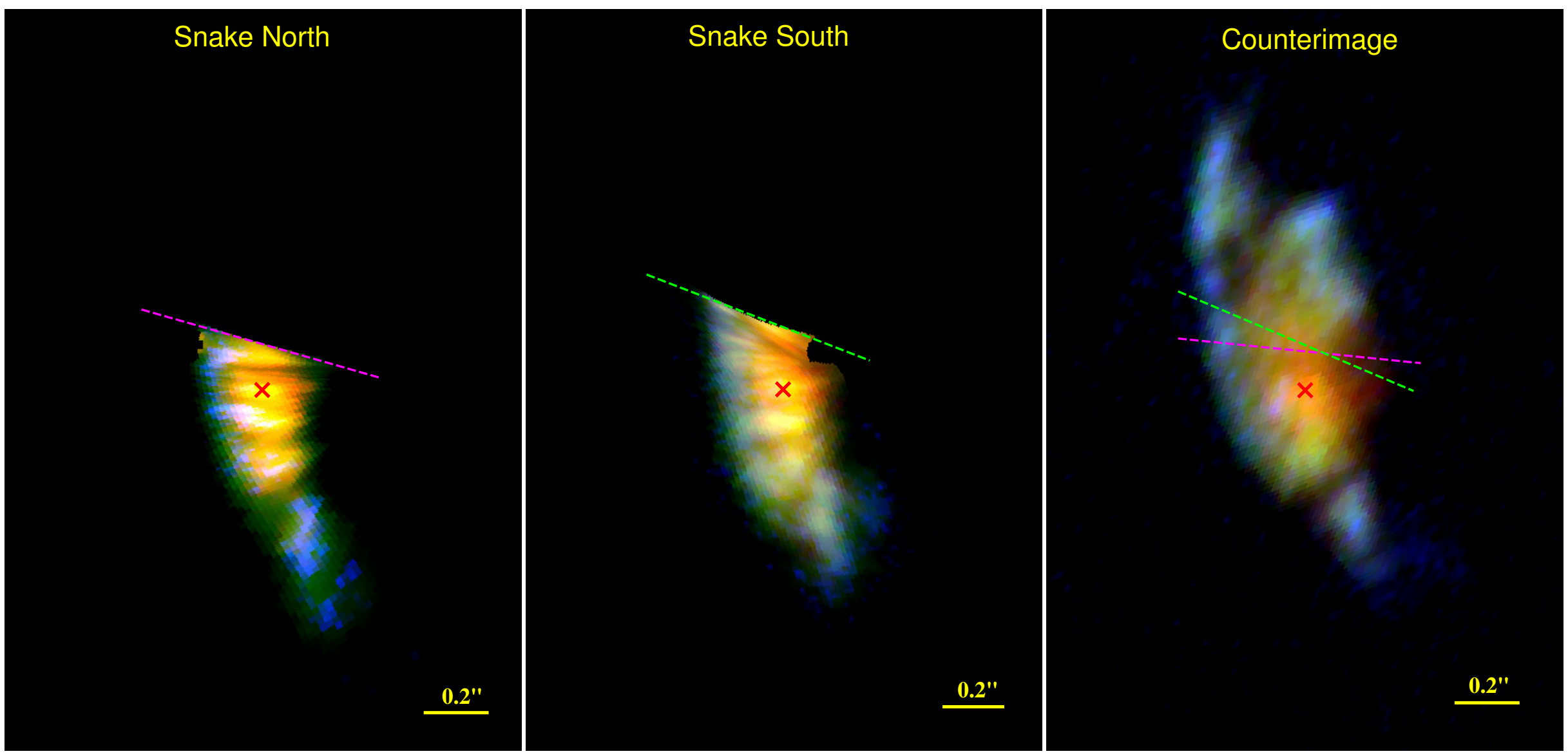

**Supplementary Figure 9** | RGB reconstruction of the source plane for the Cosmic Snake North, South, and the Counterimage (left, central, right panel respectively). The red cross marks the center of the galaxy (identified as the peak in the F160W emission) in each panel. Note that the Cosmic Snake span only about half of the source galaxy, while the Counterimage entirely reproduces the source. A reference scale bar is overplotted in each panel.

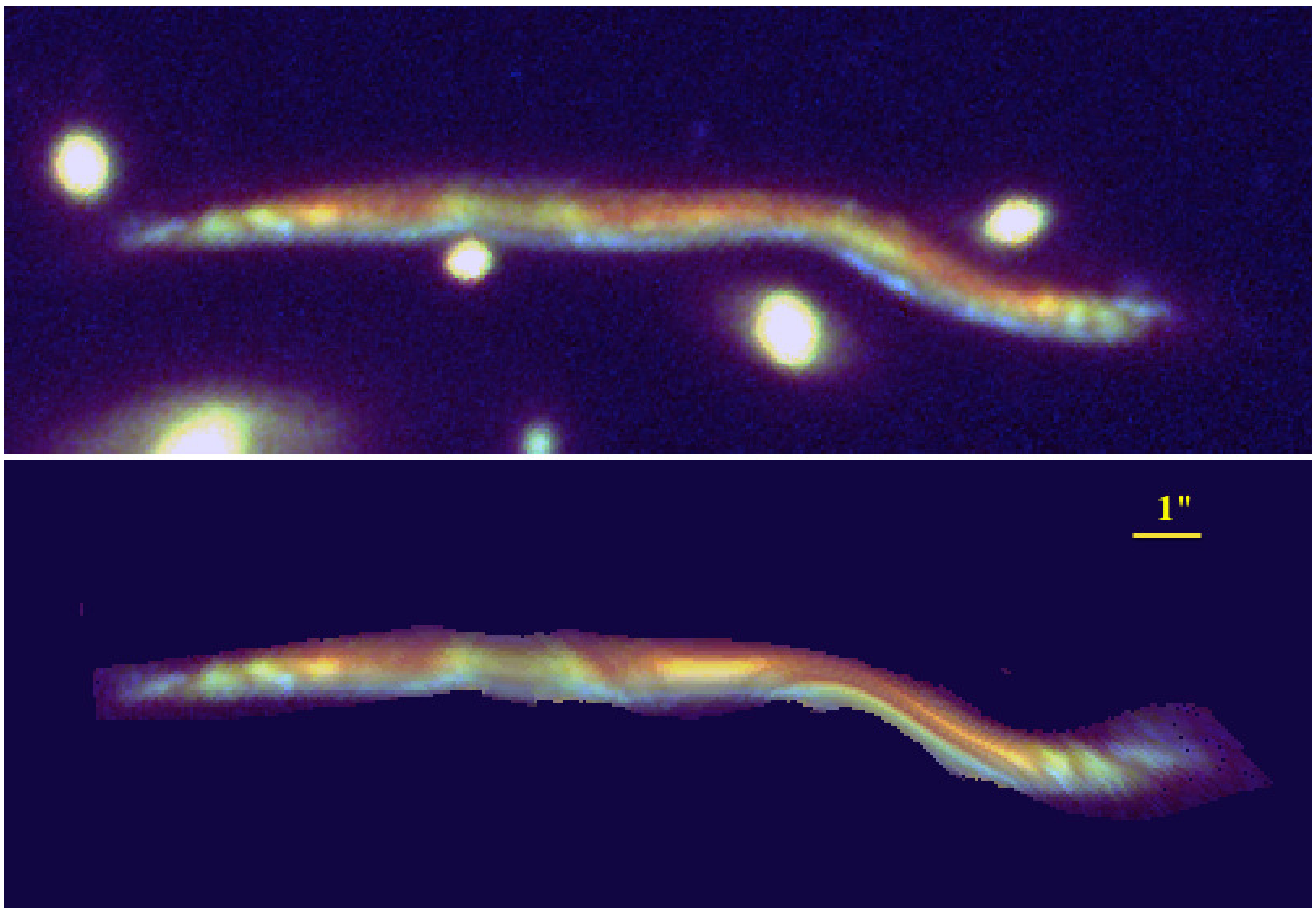

**Supplementary Figure 10** | Comparison between the observed (top panel) and simulated (bottom panel) HST color images (Blue = F606W, Green=F110W, Red=F160W) of the Cosmic Snake. The simulation is performed by reconstructing the morphology of the Northern part of the arc (left in this image) to the source plane, then re-lensing back into the image plane. A reference scale bar is overplotted in the bottom panel.

**Supplementary Table 1**. Part I. Photometric catalogue.

| ID | F275W | F336W | F390W | F435W | F475W | F606W | F625W |
|---|---|---|---|---|---|---|---|
| | WFC3 UVIS | WFC3 UVIS | WFC3 UVIS | ACS WFC | ACS WFC | ACS WFC | ACS WFC |
| CI-all | 23.10±0.03 | 22.50±0.02 | 22.35±0.01 | 22.39±0.01 | 22.23±0.01 | 21.72±0.01 | 21.57±0.01 |
| CI-R1 | — | 27.53±0.20 | 26.74±0.04 | 27.07±0.11 | 26.87±0.09 | 26.19±0.05 | 25.92±0.05 |
| CI-R2 | — | — | 26.84±0.08 | 27.01±0.14 | 26.62±0.06 | 25.95±0.04 | 25.85±0.03 |
| CI-a | — | 27.66±0.23 | 27.08±0.06 | 27.37±0.15 | 27.40±0.16 | 26.61±0.05 | 26.54±0.06 |
| CI-b | 26.79±0.09 | 26.41±0.05 | 25.85±0.02 | 25.50±0.03 | 25.33±0.02 | 25.01±0.02 | 24.93±0.02 |
| CI-c | 26.36±0.09 | 26.45±0.06 | 25.60±0.02 | 25.25±0.03 | 25.16±0.02 | 24.85±0.01 | 24.83±0.02 |
| CI-d | — | 26.78±0.07 | 26.18±0.03 | 25.74±0.04 | 25.88±0.04 | 25.51±0.03 | 25.37±0.03 |
| CI-e | 26.88±0.12 | 26.25±0.04 | 25.92±0.03 | 25.75±0.03 | 25.51±0.02 | 25.19±0.02 | 25.11±0.03 |
| CI-f | 27.18±0.13 | 27.10±0.12 | 26.51±0.06 | 26.36±0.06 | 26.27±0.05 | 25.75±0.05 | 25.64±0.05 |
| CI-g-R3 | 27.01±0.14 | 26.98±0.11 | 26.13±0.03 | 26.19±0.06 | 25.80±0.03 | 25.31±0.03 | 25.28±0.02 |
| CI-h | 27.34±0.18 | 27.21±0.11 | 26.84±0.05 | 27.13±0.11 | 26.65±0.07 | 26.15±0.03 | 26.13±0.10 |
| CI-i | 27.15±0.12 | 27.44±0.14 | 26.49±0.04 | 26.54±0.12 | 26.25±0.04 | 25.66±0.03 | 25.61±0.03 |
| CI-j | — | — | 26.85±0.08 | 26.88±0.11 | 26.51±0.06 | 25.85±0.04 | 25.90±0.04 |
| CI-k | — | — | 26.91±0.07 | 27.22±0.13 | 26.67±0.07 | 26.07±0.04 | 26.09±0.07 |
| CI-l | 27.54±0.20 | 26.71±0.08 | 25.86±0.02 | 25.82±0.04 | 25.61±0.03 | 25.05±0.02 | 25.04±0.02 |
| CI-m | — | 27.14±0.11 | 26.48±0.04 | 26.37±0.06 | 26.30±0.07 | 25.72±0.03 | 25.73±0.04 |
| CI-n | — | 27.65±0.18 | 26.73±0.05 | 26.62±0.11 | 26.64±0.06 | 26.03±0.04 | 25.95±0.05 |
| CI-o | 27.51±0.20 | 27.53±0.15 | 26.92±0.06 | 27.02±0.13 | 26.57±0.08 | 26.09±0.05 | 25.94±0.05 |
| CI-p | 27.84±0.24 | — | 27.44±0.08 | — | 27.71±0.17 | 26.92±0.07 | 26.72±0.08 |
| CI-q | 27.82±0.23 | 27.60±0.18 | 27.11±0.06 | 27.30±0.13 | 26.83±0.07 | 26.30±0.05 | 26.24±0.05 |
| CI-r | — | 27.48±0.13 | 26.92±0.06 | 27.32±0.16 | 26.79±0.08 | 26.34±0.08 | 26.25±0.07 |
| CI-s | 27.31±0.15 | — | 26.56±0.04 | 26.44±0.07 | 26.12±0.04 | 25.48±0.03 | 25.33±0.03 |
| CI-t | 27.30±0.21 | 27.83±0.20 | 26.64±0.05 | 26.39±0.07 | 26.38±0.05 | 25.58±0.04 | 25.52±0.03 |
| CI-u | 27.53±0.20 | 26.79±0.08 | 26.22±0.03 | 26.00±0.05 | 25.98±0.04 | 25.33±0.04 | 25.32±0.04 |
| CI-v | 27.27±0.20 | 26.02±0.04 | 25.78±0.02 | 25.29±0.02 | 25.23±0.02 | 24.75±0.02 | 24.80±0.03 |
| Sn-R1 | 27.60±0.18 | 27.32±0.12 | 26.51±0.03 | 26.20±0.05 | 26.10±0.04 | 25.49±0.02 | 25.21±0.03 |
| Sn-R2 | 27.48±0.20 | 27.49±0.15 | 26.49±0.03 | 26.59±0.07 | 26.23±0.04 | 25.67±0.03 | 25.43±0.04 |
| Sn-R3 | — | 27.59±0.17 | 26.26±0.03 | 26.53±0.09 | 26.14±0.05 | 25.43±0.02 | 25.13±0.03 |
| Sn-R4 | — | 27.73±0.20 | 26.46±0.03 | 26.59±0.10 | 26.56±0.06 | 25.84±0.04 | 25.71±0.04 |

**Supplementary Table 1**

| ID | F275W | F336W | F390W | F435W | F475W | F606W | F625W |
|---|---|---|---|---|---|---|---|
| Sn-R5 | — | 26.87±0.12 | 26.01±0.03 | 25.70±0.04 | 25.79±0.04 | 25.02±0.02 | 24.88±0.02 |
| Sn-a | 27.75±0.23 | 27.13±0.10 | 26.66±0.04 | 26.76±0.08 | 26.29±0.05 | 25.74±0.03 | 25.75±0.04 |
| Sn-b | — | 27.75±0.20 | 26.87±0.05 | 27.21±0.12 | 26.63±0.08 | 26.15±0.04 | 26.17±0.06 |
| Sn-c | — | 26.96±0.10 | 26.33±0.03 | 26.39±0.06 | 26.17±0.04 | 25.53±0.02 | 25.46±0.03 |
| Sn-d | 27.72±0.22 | 27.23±0.14 | 26.63±0.04 | 26.64±0.09 | 26.53±0.06 | 25.76±0.03 | 25.73±0.04 |
| Sn-e | — | 26.89±0.10 | 26.91±0.05 | 26.76±0.08 | 26.48±0.07 | 25.83±0.03 | 25.75±0.04 |
| Sn-f | — | 27.10±0.12 | 26.66±0.04 | 26.90±0.09 | 26.31±0.05 | 25.83±0.03 | 25.52±0.03 |
| Sn-g | 26.84±0.10 | 26.54±0.08 | 25.54±0.02 | 25.45±0.03 | 25.24±0.02 | 24.85±0.02 | 24.83±0.02 |
| Sn-h | 27.72±0.23 | 27.17±0.10 | 26.56±0.04 | 26.45±0.07 | 26.51±0.05 | 25.87±0.03 | 25.89±0.06 |
| Sn-i | — | 27.00±0.09 | 26.37±0.03 | 26.40±0.06 | 26.31±0.06 | 25.77±0.03 | 25.69±0.03 |
| Sn-j | 26.99±0.11 | 26.61±0.06 | 25.97±0.02 | 25.82±0.05 | 25.83±0.03 | 25.42±0.02 | 25.32±0.03 |
| Sn-k | 26.88±0.11 | 26.77±0.10 | 25.87±0.03 | 25.87±0.04 | 25.56±0.03 | 25.33±0.02 | 25.00±0.02 |
| Sn-l | — | 27.28±0.13 | 26.28±0.04 | 25.93±0.04 | 25.83±0.04 | 25.38±0.02 | 25.36±0.03 |
| Sn-m | — | 27.08±0.13 | 26.76±0.08 | 26.59±0.09 | 26.38±0.07 | 25.84±0.04 | 25.69±0.04 |
| Sn-n | 27.57±0.23 | 27.10±0.13 | 26.46±0.04 | 26.34±0.06 | 26.12±0.04 | 25.76±0.04 | 25.71±0.05 |
| Ss-R1 | 27.88±0.24 | — | 26.84±0.06 | 27.57±0.21 | 27.18±0.15 | 26.01±0.10 | 25.85±0.13 |
| Ss-R2 | 27.40±0.18 | 27.82±0.00 | 26.94±0.07 | 27.50±0.18 | 27.09±0.13 | 26.14±0.13 | 25.78±0.12 |
| Ss-R3 | — | 26.82±0.04 | 26.70±0.04 | 26.35±0.07 | 26.25±0.07 | 25.72±0.02 | 25.59±0.03 |
| Ss-R4 | — | 27.39±0.19 | 26.61±0.06 | 27.05±0.13 | 26.93±0.09 | 26.03±0.10 | 25.87±0.12 |
| Ss-R5 | 27.16±0.17 | 26.88±0.11 | 26.46±0.05 | 27.33±0.23 | 26.21±0.09 | 25.79±0.07 | 25.54±0.07 |
| Ss-R6 | 26.98±0.13 | 26.71±0.07 | 26.23±0.03 | 26.18±0.06 | 25.96±0.04 | 25.37±0.03 | 25.19±0.04 |
| Ss-R7 | 27.19±0.13 | 27.51±0.16 | 26.55±0.04 | 26.71±0.08 | 26.22±0.07 | 25.80±0.05 | 25.61±0.06 |
| Ss-a | 26.46±0.08 | 26.18±0.08 | 25.81±0.03 | 25.59±0.04 | 25.53±0.02 | 25.22±0.05 | 25.09±0.06 |
| Ss-b | 26.26±0.06 | 26.21±0.09 | 25.69±0.03 | 25.51±0.04 | 25.41±0.04 | 25.07±0.05 | 25.01±0.08 |
| Ss-c | 27.47±0.17 | 26.46±0.03 | 26.62±0.04 | 26.52±0.07 | 26.23±0.05 | 25.72±0.02 | 25.78±0.04 |
| Ss-d | 26.99±0.12 | 26.16±0.03 | 26.10±0.03 | 25.72±0.04 | 25.49±0.03 | 25.10±0.02 | 25.10±0.02 |
| Ss-e | 26.59±0.11 | 25.84±0.07 | 25.54±0.03 | 25.41±0.03 | 25.29±0.02 | 24.85±0.02 | 24.84±0.04 |
| Ss-f | 27.15±0.16 | 26.31±0.10 | 26.19±0.03 | 26.13±0.05 | 25.72±0.04 | 25.43±0.02 | 25.37±0.05 |
| Ss-g | 27.26±0.19 | 26.73±0.09 | 26.25±0.03 | 26.05±0.06 | 25.84±0.04 | 25.30±0.04 | 25.31±0.04 |
| Ss-h | 27.15±0.20 | 26.54±0.09 | 26.44±0.06 | 26.21±0.07 | 25.51±0.03 | 25.71±0.05 | 25.70±0.06 |
| Ss-i | 27.82±0.24 | 26.94±0.04 | 26.44±0.03 | 26.40±0.06 | 25.88±0.03 | 25.54±0.02 | 25.48±0.03 |

## Supplementary Table 1

| ID | F275W | F336W | F390W | F435W | F475W | F606W | F625W |
|---|---|---|---|---|---|---|---|
| Ss-j | — | 27.27±0.05 | 27.07±0.06 | 27.43±0.17 | 26.87±0.09 | 26.70±0.05 | 26.76±0.08 |
| Ss-k | 27.10±0.17 | 26.70±0.03 | 26.16±0.03 | 25.99±0.06 | 25.81±0.03 | 25.32±0.02 | 25.32±0.04 |
| Ss-l | 27.30±0.17 | 27.09±0.07 | 26.61±0.04 | 26.49±0.08 | 26.32±0.06 | 25.82±0.03 | 25.79±0.03 |
| Cn-R1 | 27.76±0.24 | 26.92±0.10 | 26.45±0.04 | 27.10±0.12 | 26.55±0.09 | 25.74±0.04 | 25.56±0.04 |
| Cn-R2 | 27.28±0.19 | — | 26.84±0.07 | 26.80±0.09 | 26.75±0.09 | 26.10±0.08 | 25.95±0.09 |
| Cn-R3 | 27.37±0.21 | — | 26.93±0.05 | 26.85±0.10 | 26.55±0.08 | 25.88±0.05 | 25.64±0.06 |
| Cn-a | — | 27.23±0.16 | 26.77±0.04 | 26.52±0.07 | 26.33±0.05 | 25.97±0.03 | 25.99±0.10 |
| Cn-b | — | 27.04±0.13 | 26.64±0.04 | 26.47±0.09 | 26.31±0.07 | 26.02±0.07 | 26.00±0.05 |
| Cn-c | 27.23±0.16 | 26.50±0.06 | 25.97±0.03 | 25.59±0.03 | 25.48±0.03 | 25.10±0.02 | 25.11±0.02 |
| Cn-d | 26.21±0.07 | 26.39±0.07 | 25.60±0.03 | 25.28±0.03 | 25.26±0.02 | 24.92±0.03 | 24.78±0.02 |
| Cn-e | 27.09±0.15 | 26.97±0.11 | 26.43±0.05 | 26.18±0.05 | 26.02±0.04 | 25.69±0.05 | 25.69±0.07 |
| Cn-f | 26.67±0.09 | 26.82±0.07 | 26.47±0.03 | 26.37±0.07 | 26.17±0.04 | 25.79±0.06 | 25.68±0.05 |
| Cn-g | 26.99±0.11 | 27.31±0.13 | 26.65±0.06 | 26.68±0.09 | 26.41±0.06 | 25.96±0.07 | 25.90±0.05 |
| Cs-R1 | — | — | 27.22±0.07 | 27.00±0.12 | 26.99±0.10 | 26.53±0.05 | 26.02±0.06 |
| Cs-R2 | — | — | 27.22±0.12 | — | 27.47±0.18 | 26.28±0.12 | 26.24±0.17 |
| Cs-R3 | — | — | 26.91±0.11 | 27.44±0.18 | 26.74±0.10 | 26.16±0.08 | 25.90±0.15 |
| Cs-a | 27.14±0.17 | 26.98±0.09 | 25.93±0.03 | 25.95±0.05 | 25.74±0.03 | 25.35±0.02 | 25.21±0.02 |
| Cs-b | — | 26.69±0.07 | 26.07±0.03 | 25.91±0.04 | 25.77±0.04 | 25.29±0.02 | 25.16±0.02 |
| Cs-c | — | 26.41±0.08 | 25.91±0.03 | 25.89±0.04 | 25.52±0.04 | 25.21±0.05 | 25.10±0.06 |
| Cs-d | 26.76±0.09 | 26.37±0.08 | 25.77±0.02 | 25.75±0.04 | 25.72±0.03 | 25.17±0.03 | 25.11±0.06 |

Note. — The photometry is provided in ABmag system. The listed magnitudes have been corrected for aperture loss and foreground extinction.

**Supplementary Table 2**. PartII. Photometric catalogue.

| ID | F775W | F814W | F850LP | F105W | F110W | F125W | F140W | F160W |
|---|---|---|---|---|---|---|---|---|
| | ACS WFC | ACS WFC | ACS WFC | WFC3 IR | WFC3 IR | WFC3 IR | WFC3 IR | WFC3 IR |
| CI-all | 20.89±0.01 | 20.72±0.01 | 20.40±0.01 | 20.18±0.01 | 19.98±0.01 | 19.85±0.01 | 19.73±0.01 | 19.64±0.01 |
| CI-R1 | 25.15±0.05 | 24.92±0.04 | 24.43±0.03 | 24.05±0.01 | 23.77±0.03 | 23.56±0.03 | 23.32±0.02 | 23.11±0.02 |
| CI-R2 | 25.00±0.02 | 24.75±0.02 | 24.38±0.02 | 24.06±0.03 | 23.83±0.03 | 23.64±0.03 | 23.49±0.03 | 23.34±0.02 |
| CI-a | 25.85±0.04 | 25.62±0.04 | 25.35±0.06 | 25.06±0.02 | 24.93±0.02 | 24.84±0.02 | 24.78±0.02 | 24.77±0.03 |
| CI-b | 24.58±0.02 | 24.37±0.01 | 24.20±0.02 | 24.11±0.01 | 23.96±0.01 | 23.89±0.01 | 23.87±0.01 | 23.85±0.01 |
| CI-c | 24.33±0.02 | 24.19±0.02 | 24.08±0.02 | 23.87±0.02 | 23.76±0.02 | 23.66±0.01 | 23.64±0.02 | 23.59±0.01 |
| CI-d | 24.96±0.05 | 24.89±0.04 | 24.70±0.06 | 24.51±0.05 | 24.35±0.05 | 24.26±0.04 | 24.19±0.05 | 24.16±0.05 |
| CI-e | 24.56±0.03 | 24.44±0.02 | 24.24±0.05 | 24.08±0.03 | 23.94±0.03 | 23.85±0.05 | 23.77±0.05 | 23.71±0.03 |
| CI-f | 25.09±0.09 | 24.93±0.08 | 24.60±0.07 | 24.47±0.09 | 24.32±0.10 | 24.24±0.10 | 24.13±0.02 | 24.02±0.10 |
| CI-g-R3 | 24.63±0.04 | 24.50±0.03 | 24.24±0.03 | 24.03±0.04 | 23.79±0.03 | 23.63±0.01 | 23.46±0.03 | 23.36±0.03 |
| CI-h | 25.46±0.11 | 25.29±0.08 | 25.01±0.14 | 24.81±0.15 | 24.71±0.16 | 24.61±0.17 | 24.54±0.17 | 24.42±0.16 |
| CI-i | 24.90±0.04 | 24.76±0.02 | 24.40±0.03 | 24.23±0.05 | 24.07±0.05 | 23.92±0.05 | 23.82±0.05 | 23.71±0.05 |
| CI-j | 24.97±0.03 | 24.80±0.02 | 24.39±0.04 | 24.14±0.03 | 23.99±0.04 | 23.86±0.03 | 23.74±0.03 | 23.60±0.02 |
| CI-k | 25.16±0.03 | 24.92±0.04 | 24.58±0.04 | 24.27±0.02 | 24.10±0.03 | 23.94±0.02 | 23.81±0.02 | 23.69±0.03 |
| CI-l | 24.52±0.03 | 24.40±0.02 | 24.16±0.03 | 24.04±0.02 | 23.88±0.03 | 23.75±0.03 | 23.68±0.03 | 23.61±0.03 |
| CI-m | 25.19±0.05 | 25.03±0.04 | 24.74±0.05 | 24.68±0.04 | 24.51±0.05 | 24.40±0.03 | 24.31±0.05 | 24.24±0.04 |
| CI-n | 25.49±0.06 | 25.25±0.05 | 24.97±0.05 | 24.91±0.05 | 24.73±0.04 | 24.60±0.05 | 24.53±0.05 | 24.48±0.05 |
| CI-o | 25.48±0.06 | 25.38±0.06 | 25.15±0.06 | 25.03±0.05 | 24.87±0.05 | 24.75±0.06 | 24.65±0.05 | 24.64±0.05 |
| CI-p | 26.26±0.07 | 26.07±0.06 | 25.79±0.07 | 25.67±0.04 | 25.47±0.03 | 25.32±0.03 | 25.18±0.03 | 25.28±0.04 |
| CI-q | 25.80±0.04 | 25.63±0.03 | 25.36±0.04 | 25.30±0.03 | 25.09±0.02 | 24.95±0.03 | 24.92±0.02 | 24.98±0.03 |
| CI-r | 25.70±0.09 | 25.60±0.07 | 25.32±0.06 | 25.23±0.05 | 25.06±0.06 | 24.91±0.06 | 24.82±0.03 | 24.82±0.05 |
| CI-s | 24.85±0.04 | 24.63±0.03 | 24.31±0.02 | 24.16±0.04 | 23.96±0.03 | 23.85±0.03 | 23.74±0.04 | 23.64±0.04 |
| CI-t | 25.00±0.05 | 24.78±0.02 | 24.47±0.02 | 24.31±0.05 | 24.15±0.05 | 24.02±0.05 | 23.94±0.02 | 23.84±0.05 |
| CI-u | 24.79±0.05 | 24.60±0.03 | 24.23±0.04 | 24.14±0.04 | 24.00±0.04 | 23.90±0.01 | 23.83±0.04 | 23.74±0.02 |
| CI-v | 24.32±0.01 | 24.14±0.02 | 23.86±0.03 | 23.77±0.03 | 23.64±0.02 | 23.56±0.01 | 23.49±0.03 | 23.43±0.01 |
| Sn-R1 | 24.57±0.03 | 24.39±0.02 | 24.01±0.02 | 23.80±0.01 | 23.62±0.01 | 23.49±0.01 | 23.37±0.02 | 23.24±0.01 |
| Sn-R2 | 24.62±0.02 | 24.39±0.02 | 24.02±0.02 | 23.69±0.02 | 23.47±0.02 | 23.28±0.02 | 23.14±0.01 | 22.99±0.02 |
| Sn-R3 | 24.28±0.02 | 24.14±0.02 | 23.67±0.02 | 23.26±0.02 | 22.93±0.01 | 22.69±0.01 | 22.44±0.01 | 22.23±0.01 |
| Sn-R4 | 24.91±0.03 | 24.69±0.02 | 24.33±0.03 | 24.00±0.02 | 23.75±0.01 | 23.59±0.02 | 23.38±0.01 | 23.19±0.01 |

## Supplementary Table 2

| ID | F775W | F814W | F850LP | F105W | F110W | F125W | F140W | F160W |
|---|---|---|---|---|---|---|---|---|
| Sn-R5 | 24.13±0.02 | 23.88±0.01 | 23.49±0.01 | 23.25±0.02 | 23.04±0.02 | 22.90±0.02 | 22.75±0.02 | 22.63±0.02 |
| Sn-a | 25.03±0.03 | 24.99±0.02 | 24.81±0.03 | 24.62±0.02 | 24.43±0.02 | 24.26±0.02 | 24.19±0.02 | 24.16±0.02 |
| Sn-b | 25.51±0.04 | 25.34±0.03 | 25.18±0.04 | 24.98±0.03 | 24.85±0.03 | 24.71±0.03 | 24.62±0.03 | 24.59±0.03 |
| Sn-c | 24.93±0.02 | 24.78±0.02 | 24.52±0.02 | 24.40±0.02 | 24.24±0.02 | 24.12±0.02 | 24.05±0.02 | 23.94±0.02 |
| Sn-d | 25.10±0.03 | 24.98±0.02 | 24.77±0.03 | 24.62±0.02 | 24.40±0.02 | 24.27±0.02 | 24.19±0.02 | 24.12±0.02 |
| Sn-e | 25.06±0.03 | 24.93±0.02 | 24.53±0.03 | 24.38±0.02 | 24.26±0.02 | 24.12±0.02 | 24.04±0.02 | 23.96±0.02 |
| Sn-f | 24.98±0.03 | 24.83±0.02 | 24.49±0.02 | 24.39±0.02 | 24.24±0.02 | 24.13±0.02 | 24.03±0.02 | 23.93±0.02 |
| Sn-g | 24.22±0.02 | 24.06±0.01 | 23.74±0.02 | 23.58±0.02 | 23.37±0.02 | 23.23±0.02 | 23.12±0.02 | 23.00±0.02 |
| Sn-h | 25.15±0.04 | 25.04±0.03 | 24.70±0.03 | 24.51±0.02 | 24.38±0.02 | 24.23±0.02 | 24.13±0.02 | 23.99±0.02 |
| Sn-i | 25.09±0.03 | 24.87±0.02 | 24.53±0.03 | 24.43±0.02 | 24.26±0.02 | 24.17±0.03 | 24.03±0.02 | 23.95±0.02 |
| Sn-j | 24.71±0.02 | 24.58±0.02 | 24.34±0.02 | 24.19±0.02 | 24.03±0.02 | 23.97±0.02 | 23.85±0.02 | 23.77±0.02 |
| Sn-k | 24.60±0.02 | 24.41±0.02 | 24.13±0.03 | 23.94±0.02 | 23.79±0.03 | 23.69±0.02 | 23.58±0.02 | 23.44±0.02 |
| Sn-l | 24.78±0.02 | 24.59±0.02 | 24.28±0.03 | 24.20±0.02 | 24.02±0.03 | 23.92±0.02 | 23.81±0.02 | 23.73±0.02 |
| Sn-m | 25.04±0.04 | 24.87±0.02 | 24.62±0.04 | 24.41±0.03 | 24.25±0.03 | 24.16±0.03 | 24.04±0.03 | 23.98±0.03 |
| Sn-n | 25.00±0.03 | 24.80±0.04 | 24.49±0.04 | 24.38±0.04 | 24.20±0.02 | 24.09±0.02 | 23.97±0.02 | 23.88±0.05 |
| Ss-R1 | 24.99±0.12 | 24.75±0.11 | 24.26±0.11 | 23.98±0.09 | 23.75±0.10 | 23.56±0.08 | 23.42±0.08 | 23.27±0.09 |
| Ss-R2 | 24.98±0.10 | 24.79±0.10 | 24.40±0.11 | 24.01±0.09 | 23.76±0.08 | 23.53±0.07 | 23.37±0.07 | 23.24±0.07 |
| Ss-R3 | 24.96±0.02 | 24.85±0.02 | 24.52±0.02 | 24.27±0.02 | 24.06±0.01 | 23.92±0.01 | 23.76±0.01 | 23.62±0.01 |
| Ss-R4 | 25.01±0.03 | 24.78±0.02 | 24.33±0.10 | 23.95±0.01 | 23.69±0.02 | 23.49±0.08 | 23.24±0.07 | 23.02±0.04 |
| Ss-R5 | 24.61±0.06 | 24.39±0.07 | 23.89±0.02 | 23.48±0.05 | 23.16±0.04 | 22.91±0.01 | 22.67±0.04 | 22.46±0.03 |
| Ss-R6 | 24.44±0.02 | 24.32±0.03 | 23.99±0.02 | 23.68±0.03 | 23.46±0.03 | 23.31±0.03 | 23.16±0.01 | 23.05±0.01 |
| Ss-R7 | 24.85±0.03 | 24.69±0.02 | 24.28±0.02 | 24.10±0.02 | 23.90±0.03 | 23.77±0.03 | 23.63±0.03 | 23.53±0.03 |
| Ss-a | 24.65±0.08 | 24.56±0.09 | 24.36±0.09 | 24.31±0.12 | 24.10±0.12 | 23.96±0.12 | 23.89±0.13 | 23.94±0.15 |
| Ss-b | 24.46±0.09 | 24.33±0.08 | 24.16±0.09 | 24.04±0.11 | 23.82±0.13 | 23.68±0.10 | 23.60±0.11 | 23.58±0.12 |
| Ss-c | 25.09±0.03 | 24.96±0.02 | 24.59±0.02 | 24.53±0.02 | 24.36±0.02 | 24.25±0.02 | 24.11±0.02 | 24.06±0.02 |
| Ss-d | 24.53±0.02 | 24.42±0.01 | 24.15±0.02 | 23.99±0.01 | 23.81±0.01 | 23.69±0.01 | 23.57±0.01 | 23.50±0.01 |
| Ss-e | 24.19±0.05 | 24.07±0.01 | 23.72±0.05 | 23.55±0.06 | 23.36±0.06 | 23.23±0.06 | 23.06±0.01 | 22.95±0.06 |
| Ss-f | 24.75±0.06 | 24.61±0.07 | 24.33±0.08 | 24.21±0.09 | 24.06±0.01 | 23.92±0.10 | 23.83±0.10 | 23.71±0.09 |
| Ss-g | 24.59±0.04 | 24.53±0.04 | 24.24±0.03 | 24.01±0.02 | 23.82±0.02 | 23.70±0.04 | 23.55±0.05 | 23.48±0.04 |
| Ss-h | 25.16±0.06 | 24.94±0.07 | 24.59±0.06 | 24.51±0.09 | 24.40±0.09 | 24.28±0.09 | 24.21±0.10 | 24.13±0.10 |
| Ss-i | 24.73±0.02 | 24.68±0.01 | 24.28±0.02 | 24.17±0.01 | 24.03±0.01 | 23.91±0.01 | 23.81±0.01 | 23.71±0.01 |

**Supplementary Table 2**

| ID | F775W | F814W | F850LP | F105W | F110W | F125W | F140W | F160W |
|---|---|---|---|---|---|---|---|---|
| Ss-j | 26.22±0.08 | 25.90±0.04 | 25.69±0.06 | 25.48±0.04 | 25.32±0.03 | 25.19±0.03 | 25.05±0.03 | 25.02±0.03 |
| Ss-k | 24.83±0.03 | 24.59±0.01 | 24.34±0.02 | 24.25±0.01 | 24.08±0.01 | 23.96±0.01 | 23.90±0.01 | 23.82±0.01 |
| Ss-l | 25.29±0.03 | 25.17±0.02 | 24.88±0.03 | 24.84±0.02 | 24.63±0.02 | 24.54±0.02 | 24.46±0.02 | 24.40±0.02 |
| Cn-R1 | 24.69±0.03 | 24.39±0.02 | 23.99±0.02 | 23.63±0.02 | 23.40±0.01 | 23.22±0.02 | 23.06±0.02 | 22.91±0.02 |
| Cn-R2 | 25.09±0.06 | 24.89±0.06 | 24.46±0.05 | 24.14±0.05 | 23.88±0.01 | 23.68±0.05 | 23.47±0.04 | 23.26±0.04 |
| Cn-R3 | 24.85±0.07 | 24.62±0.07 | 24.18±0.06 | 23.87±0.02 | 23.61±0.01 | 23.36±0.05 | 23.17±0.05 | 22.98±0.05 |
| Cn-a | 25.26±0.09 | 25.14±0.07 | 24.79±0.04 | 24.72±0.03 | 24.54±0.09 | 24.43±0.09 | 24.36±0.03 | 24.29±0.03 |
| Cn-b | 25.32±0.04 | 25.14±0.09 | 24.94±0.12 | 24.77±0.03 | 24.68±0.04 | 24.52±0.13 | 24.46±0.03 | 24.40±0.05 |
| Cn-c | 24.62±0.02 | 24.42±0.02 | 24.21±0.03 | 24.10±0.04 | 23.94±0.02 | 23.81±0.04 | 23.71±0.02 | 23.66±0.02 |
| Cn-d | 24.32±0.03 | 24.13±0.03 | 23.86±0.03 | 23.71±0.03 | 23.52±0.03 | 23.45±0.02 | 23.29±0.03 | 23.18±0.04 |
| Cn-e | 25.18±0.08 | 24.99±0.07 | 24.69±0.08 | 24.57±0.10 | 24.45±0.08 | 24.36±0.10 | 24.30±0.03 | 24.16±0.10 |
| Cn-f | 25.16±0.07 | 24.96±0.07 | 24.71±0.08 | 24.51±0.10 | 24.43±0.10 | 24.28±0.10 | 24.19±0.11 | 24.10±0.13 |
| Cn-g | 25.24±0.09 | 25.00±0.10 | 24.73±0.09 | 24.54±0.13 | 24.43±0.13 | 24.30±0.11 | 24.16±0.11 | 24.06±0.14 |
| Cs-R1 | 25.35±0.04 | 25.16±0.02 | 24.78±0.03 | 24.42±0.02 | 24.17±0.02 | 23.98±0.02 | 23.79±0.01 | 23.64±0.01 |
| Cs-R2 | 25.23±0.14 | 25.04±0.07 | 24.59±0.12 | 24.23±0.16 | 24.00±0.16 | 23.81±0.13 | 23.64±0.11 | 23.49±0.11 |
| Cs-R3 | 25.02±0.08 | 24.85±0.13 | 24.37±0.11 | 24.10±0.12 | 23.87±0.12 | 23.70±0.11 | 23.55±0.10 | 23.39±0.10 |
| Cs-a | 24.56±0.02 | 24.48±0.02 | 24.19±0.02 | 23.94±0.01 | 23.73±0.02 | 23.61±0.01 | 23.47±0.01 | 23.32±0.01 |
| Cs-b | 24.61±0.02 | 24.48±0.02 | 24.22±0.02 | 23.99±0.01 | 23.81±0.01 | 23.67±0.01 | 23.53±0.01 | 23.47±0.01 |
| Cs-c | 24.61±0.05 | 24.57±0.09 | 24.34±0.08 | 24.26±0.12 | 24.06±0.13 | 23.91±0.14 | 23.85±0.14 | 23.84±0.14 |
| Cs-d | 24.65±0.08 | 24.57±0.09 | 24.36±0.09 | 24.21±0.12 | 24.03±0.13 | 23.87±0.13 | 23.82±0.13 | 23.83±0.15 |

Note. — The photometry is provided in ABmag system. The listed magnitudes have been corrected for aperture loss and foreground extinction.

**Supplementary Table 3**. Structural and physical properties of the clumps.

| ID | a[a] | b[b] | $R_\mu$[c] | $\log(M_\mu$[d]$)$ | $\mu$[e] |
|---|---|---|---|---|---|
| | ″ | ″ | pc | $M_\odot$ | |
| CI-all | 1.00 | 4.20 | 7957 | 10.56 | 4.3 |
| CI-R1 | 0.09 | 0.12 | 325 | 9.51 | 4.0 |
| CI-R2 | 0.08 | 0.12 | 291 | 9.09 | 4.1 |
| CI-a | 0.08 | 0.09 | 322 | 8.46 | 3.5 |
| CI-b | 0.11 | 0.15 | 529 | 8.31 | 3.6 |
| CI-c | 0.14 | 0.17 | 613 | 8.42 | 3.6 |
| CI-d | 0.09 | 0.14 | 445 | 8.36 | 3.7 |
| CI-e | 0.11 | 0.16 | 526 | 8.73 | 3.8 |
| CI-f | 0.09 | 0.11 | 378 | 8.67 | 4.0 |
| CI-g-R3 | 0.10 | 0.13 | 427 | 9.01 | 4.0 |
| CI-h | 0.09 | 0.09 | 309 | 8.53 | 4.1 |
| CI-i | 0.09 | 0.13 | 407 | 8.88 | 4.1 |
| CI-j | 0.08 | 0.12 | 361 | 9.03 | 4.2 |
| CI-k | 0.08 | 0.12 | 358 | 8.94 | 4.2 |
| CI-l | 0.12 | 0.16 | 516 | 8.81 | 4.4 |
| CI-m | 0.09 | 0.09 | 315 | 8.48 | 4.5 |
| CI-n | 0.08 | 0.09 | 291 | 8.40 | 4.6 |
| CI-o | 0.08 | 0.09 | 289 | 8.32 | 4.6 |
| CI-p | 0.08 | 0.09 | 283 | 8.06 | 4.8 |
| CI-q | 0.09 | 0.09 | 289 | 8.17 | 4.7 |
| CI-r | 0.09 | 0.09 | 291 | 8.23 | 4.6 |
| CI-s | 0.10 | 0.12 | 425 | 8.95 | 3.9 |
| CI-t | 0.08 | 0.08 | 296 | 8.83 | 3.8 |
| CI-u | 0.09 | 0.12 | 406 | 8.81 | 3.8 |
| CI-v | 0.12 | 0.18 | 598 | 8.97 | 3.7 |
| Sn-R1 | 0.11 | 0.16 | 300 | 8.72 | 9.7 |
| Sn-R2 | 0.12 | 0.16 | 242 | 8.73 | 16.0 |
| Sn-R3 | 0.15 | 0.26 | 274 | 8.82 | 29.0 |
| Sn-R4 | 0.11 | 0.16 | 127 | 8.06 | 50.3 |
| Sn-R5 | 0.16 | 0.24 | 262 | 8.54 | 32.4 |
| Sn-a | 0.08 | 0.16 | 482 | 8.71 | 3.2 |
| Sn-b | 0.08 | 0.10 | 350 | 8.50 | 3.6 |
| Sn-c | 0.09 | 0.12 | 358 | 8.64 | 4.6 |
| Sn-d | 0.08 | 0.08 | 243 | 8.46 | 5.7 |
| Sn-e | 0.08 | 0.12 | 262 | 8.47 | 7.9 |
| Sn-f | 0.09 | 0.13 | 261 | 8.38 | 10.0 |
| Sn-g | 0.15 | 0.18 | 308 | 8.48 | 17.1 |
| Sn-h | 0.10 | 0.10 | 152 | 7.92 | 21.8 |
| Sn-i | 0.10 | 0.12 | 145 | 7.76 | 33.1 |
| Sn-j | 0.09 | 0.17 | 134 | 7.58 | 50.5 |

## Supplementary Table 3

| ID | a[a] | b[b] | $R_\mu$[c] | log($M_\mu$[d]) | $\mu$[e] |
|---|---|---|---|---|---|
| Sn-k | 0.09 | 0.21 | 110 | 7.53 | 93.7 |
| Sn-l | 0.08 | 0.12 | 116 | 7.85 | 40.4 |
| Sn-m | 0.09 | 0.09 | 135 | 7.99 | 21.2 |
| Sn-n | 0.10 | 0.14 | 77 | 7.20 | 134.7 |
| Ss-R1 | 0.10 | 0.12 | 61 | 7.71 | 136.2 |
| Ss-R2 | 0.10 | 0.13 | 89 | 8.00 | 71.2 |
| Ss-R3 | 0.09 | 0.09 | 77 | 7.77 | 42.1 |
| Ss-R4 | 0.09 | 0.16 | 150 | 8.71 | 29.1 |
| Ss-R5 | 0.14 | 0.21 | 306 | 9.22 | 17.5 |
| Ss-R6 | 0.12 | 0.15 | 270 | 8.75 | 12.3 |
| Ss-R7 | 0.09 | 0.12 | 201 | 8.63 | 10.6 |
| Ss-a | 0.09 | 0.13 | 49 | 6.35 | 261.1 |
| Ss-b | 0.09 | 0.18 | 121 | 7.55 | 66.2 |
| Ss-c | 0.09 | 0.09 | 104 | 7.56 | 41.2 |
| Ss-d | 0.09 | 0.18 | 172 | 7.87 | 33.0 |
| Ss-e | 0.11 | 0.29 | 292 | 8.24 | 23.3 |
| Ss-f | 0.10 | 0.14 | 225 | 8.11 | 16.1 |
| Ss-g | 0.09 | 0.11 | 213 | 8.39 | 12.4 |
| Ss-h | 0.10 | 0.10 | 214 | 8.00 | 11.7 |
| Ss-i | 0.09 | 0.12 | 245 | 8.43 | 10.3 |
| Ss-j | 0.08 | 0.11 | 228 | 8.00 | 9.5 |
| Ss-k | 0.09 | 0.14 | 285 | 8.34 | 9.1 |
| Ss-l | 0.08 | 0.11 | 239 | 8.09 | 8.6 |
| Cn-R1 | 0.13 | 0.19 | 168 | 8.36 | 44.7 |
| Cn-R2 | 0.09 | 0.14 | 107 | 8.29 | 47.4 |
| Cn-R3 | 0.10 | 0.18 | 44 | 7.26 | 451.1 |
| Cn-a | 0.09 | 0.09 | 87 | 7.41 | 51.3 |
| Cn-b | 0.09 | 0.09 | 94 | 7.41 | 44.0 |
| Cn-c | 0.09 | 0.16 | 147 | 7.67 | 39.3 |
| Cn-d | 0.11 | 0.26 | 219 | 7.94 | 37.0 |
| Cn-e | 0.09 | 0.09 | 99 | 7.53 | 39.5 |
| Cn-f | 0.08 | 0.12 | 105 | 7.65 | 47.1 |
| Cn-g | 0.09 | 0.09 | 71 | 7.54 | 76.9 |
| Cs-R1 | 0.08 | 0.08 | 62 | 8.01 | 36.3 |
| Cs-R2 | 0.08 | 0.10 | 47 | 7.67 | 110.2 |
| Cs-R3 | 0.08 | 0.12 | 38 | 7.42 | 231.7 |
| Cs-a | 0.10 | 0.22 | 226 | 8.18 | 26.4 |
| Cs-b | 0.10 | 0.20 | 235 | 8.19 | 22.2 |
| Cs-c | 0.09 | 0.14 | 127 | 7.52 | 45.8 |
| Cs-d | 0.10 | 0.13 | 73 | 6.92 | 140.4 |

Note. — [a]Semi-minor axis. [b]Semi-major axis. [c]Demagnified equivalent radius: $R_\mu = R_{vis}/\sqrt{\mu}$, where the circularized radius, $R_{vis}$, is defined from the semi-axes, $a$ and $b$, of the elliptical regions as: $R_{vis} = \sqrt{a \cdot b}$ [d]Demagnified mass: $M_\mu$=Mass/$\mu$. [e]Harmonic average amplification factor.

**Supplementary Table 4** | Constraints used for the lensing modelling.

| ID | $\alpha$ [hms] | $\delta$ [dms] | xID |
|---|---|---|---|
| A.1 | 12:06:10.76 | -08:47:58.02 | Sn-a |
| A.2 | 12:06:10.82 | -08:48:11.26 | Ss-l |
| B.1 | 12:06:10.76 | -08:47:58.34 | Sn-c |
| B.2 | 12:06:10.82 | -08:48:10.97 | Ss-k |
| B.3 | 12:06:11.23 | -08:47:44.45 | CI-l |
| C.1 | 12:06:10.75 | -08:47:59.14 | Sn-R1 |
| C.2 | 12:06:10.83 | -08:48:10.49 | Ss-R7 |
| D.1 | 12:06:10.74 | -08:48:00.46 | Sn-R3 |
| D.2 | 12:06:10.82 | -08:48:09.76 | Ss-R5 |
| D.3 | 12:06:11.27 | -08:47:43.59 | CI-R1 |
| E.1 | 12:06:10.74 | -08:48:02.20 | Sn-R5 |
| E.2 | 12:06:10.77 | -08:48:07.72 | Ss-R2 |
| E.4 | 12:06:10.74 | -08:48:03.65 | Cn-R1 |
| E.5 | 12:06:10.76 | -08:48:07.24 | Cs-R2 |
| F.2 | 12:06:10.80 | -08:48:07.98 | Ss-b |
| F.5 | 12:06:10.78 | -08:48:07.24 | Cs-c |
| 2.1 | 12:06:14.53 | -08:48:32.37 | |
| 2.2 | 12:06:15.00 | -08:48:17.67 | |
| 2.3 | 12:06:15.03 | -08:47:48.07 | |
| 3.1 | 12:06:14.43 | -08:48:34.20 | |
| 3.2 | 12:06:15.00 | -08:48:16.50 | |
| 3.3 | 12:06:15.01 | -08:47:48.65 | |

Note — Individual constraints used in our model for the optimisation of the cluster mass distribution. Letters A to F refer to matching groups of clumps identified in the Cosmic Snake (see Figure 8), the correspondence with the clump ID from Table 3 is provided in the last column. The last three lines refer to the images of systems 2 and 3 to the East of the cluster, spectroscopically confirmed at $z = 3.038$.

**Supplementary Table 5** | Best fit parameters of mass model.

| Potential | $\Delta\alpha$ [arcsec] | $\Delta\delta$ [arcsec] | $e$ | $\theta$ [deg] | $r_{core}$ kpc | $r_{cut}$ kpc | $\sigma$ km/s |
|---|---|---|---|---|---|---|---|
| DM1 | $0.7^{+0.4}_{-0.3}$ | $-0.8^{+0.2}_{-0.2}$ | $0.49^{+0.05}_{-0.03}$ | $4.9^{+1.0}_{-1.4}$ | $130^{+14}_{-16}$ | $2310^{+1165}_{-539}$ | $1124^{+73}_{-56}$ |
| DM2 | $[-52.0]$ | $[-19.0]$ | $0.58^{+0.04}_{-0.07}$ | $5.9^{+4.1}_{-0.2}$ | $34^{+15}_{-12}$ | $[800]$ | $322^{+47}_{-35}$ |
| BCG | $[0.0]$ | $[0.0]$ | $[0.50]$ | $[19.7]$ | $[0]$ | $250^{+13}_{-23}$ | $491^{+2}_{-29}$ |
| GAL1 | $[20.7]$ | $[-6.0]$ | $0.34^{+0.11}_{-0.18}$ | $273.0^{+12.7}_{-9.4}$ | $[0]$ | $86^{+34}_{-26}$ | $83^{+7}_{-14}$ |
| GAL2 | $[19.3]$ | $[-3.1]$ | $0.36^{+0.12}_{-0.13}$ | $53.8^{+10.2}_{-8.3}$ | $[0]$ | $108^{+89}_{-3}$ | $118^{+10}_{-5}$ |
| GAL3 | $[20.2]$ | $[1.0]$ | $[0.00]$ | $[67.3]$ | $[0]$ | $22^{+7}_{-84}$ | $51^{+7}_{-10}$ |
| GAL4 | $[21.5]$ | $[6.0]$ | $0.21^{+0.09}_{-0.11}$ | $5.7^{+12.2}_{-16.8}$ | $[0]$ | $114^{+24}_{-22}$ | $128^{+13}_{-7}$ |
| $L^*$ galaxy | | | | | $[0.15]$ | $48^{+16}_{-11}$ | $176^{+1}_{-16}$ |

Best-fit pseudo-isothermal model parameters of the cluster-scale mass profiles (Western and Eastern components), galaxy potentials modelled individually and the scaling relation of cluster members (shown for a $L^*$ galaxy). From left to right: center in arcsecs from the origin fixed at the location of the brightest cluster galaxy, ellipticity and position angle, core and cut radii, central velocity dispersion. Values in brackets are fixed in the modelling.